\documentclass[reprint,amsmath,amssymb,aps,prb,floatfix,superscriptaddress]{revtex4-2}

\usepackage[margin=1in]{geometry}
\usepackage{amsmath, amssymb}
\usepackage{physics} 
\usepackage{graphicx}

\usepackage{array}
\usepackage{multirow}
\usepackage{pifont}
\newcommand{\cmark}{\ding{52}}%
\newcommand{\xmark}{\ding{56}}%

\usepackage{hyperref}
\hypersetup{
    colorlinks=true, 
    urlcolor=blue,   
    linkcolor=blue,  
    citecolor=red    
}

\begin{document}

\title{Symmetries of Spin-Splitting Induced by Spin-Orbit Coupling in Non-magnetic Crystals}

\author{Fan Yang}
\affiliation{Department of Chemical Engineering and Materials Science, University of Minnesota, MN 55455, USA}
\author{Rafael M. Fernandes}
\affiliation{Department of Physics, The Grainger College of Engineering, University of Illinois Urbana-Champaign, Urbana, Illinois 61801, USA}
\affiliation{Anthony J. Leggett Institute for Condensed Matter Theory, The Grainger College of Engineering, University of Illinois Urbana-Champaign, Urbana, Illinois 61801, USA}
\author{Turan Birol}
\affiliation{Department of Chemical Engineering and Materials Science, University of Minnesota, MN 55455, USA}
\date{\today}

\begin{abstract}
    
Spin-orbit coupling (SOC) leads to splitting of otherwise spin-degenerate bands in noncentrosymmetric materials, even if time-reversal symmetry is present. While this gives rise to well-known phenomena such as the Rashba and Dresselhaus effects, various other terms are allowed based on the point group of the crystal and the electronic Hamiltonian. In this study, we utilize point group representations to illustrate that four different types of SOC terms (Rashba, Dresselhaus, Weyl, and Ising) can emerge in periodic solids. We construct reciprocal-space energy expressions for each type of SOC-induced splitting of opposite-spin bands, and follow a similar procedure to also obtain minimal tight-binding models that capture all types of spin-splittings for subgroups of two high-symmetry groups $m\bar{3}m$ and $6/mmm$. Furthermore, we also obtain a complete list of nodal features in the electronic band structure in these systems, distinguishing between crystallographic-symmetry-imposed nodal lines and those imposed by time-reversal-symmetry only. Finally, we conclude by presenting a list of materials that host each type of inversion-breaking SOC effects. Our classification of the spin-splitting symmetries in non-magnetic systems with SOC is the counterpart of the recent classification of spin-splitting symmetries in unconventional magnetic systems without SOC, such as altermagnets and odd-parity magnets. More broadly, our work provides a basis for studying superconductivity and other collective electronic phenomena that are impacted by SOC-induced band splittings in noncentrosymmetric materials. 
\end{abstract}

\maketitle

\section{Introduction}
\label{sec:intro}

Spin-orbit coupling (SOC) is the term of electronic Hamiltonians that couples spin degrees of freedom with spatial wavefunctions, thus explicitly breaking spin-rotational symmetry and subjecting the spins to the same symmetry operations as the crystal lattice. It gives rise to an array of unique interactions and phenomena in magnetically ordered systems, such as the Dzyaloshinskii-Moriya interaction, linear magnetoelectric effect, and the emergence of skyrmions~\cite{Dzyaloshinsky1958, Moriya1960, Birol2012Review, Nagaosa2013}. 
SOC can also fundamentally impact the electronic properties of systems with time-reversal symmetry ($\mathcal{T}$), for example, by splitting the spin-degenerate bands in noncentrosymmetric (NCS) systems, or by helping stabilize a Mott insulating phase in correlated systems~\cite{Kim2008, Birol2015}. In metals, SOC can also result in different types of superconducting phenomena by entangling spin and orbital angular momenta, such as quintet- and septet-pairing~\cite{Brydon2016,Venderbos2018, Samokhin2019Superconducting, Dutta2021} and admixtures between singlet and triplet pairing states even if spatial inversion symmetry ($\mathcal{I}$) is not broken \cite{Smidman2017,Vafek2017,Vafek2018,HYKee2021}.  

The combination of $\mathcal{T}$ and $\mathcal{I}$, referred to as the anti-inversion $\mathcal{I}'=\mathcal{TI}$, keeps opposite spin bands degenerate at all $\vb*{k}$-points (crystal momenta) if it is a symmetry of a solid. In NCS systems, however, SOC induces a splitting of otherwise spin-degenerate bands in the reciprocal space, and gives rise to a variety of spin textures that reflect the symmetry of the host crystal. The two most well-known examples of such spin-splitting patterns are those of the Rashba and Dresselhaus effects~\cite{Rashba1984, Dresselhaus1955}. The Rashba effect emerges in polar systems and gives rise to a $\sim\vb*{\sigma}\times\vb*{k}$ term in the reciprocal space Hamiltonian. Here, $\vb*{\sigma}$ is the vector of Pauli spin matrices, and the word polar refers to the crystal's symmetry, rather than the presence of a net electric dipole. The Dresselhaus effect, on the other hand, was originally introduced in cubic zinc blende structures where there is no polar axis even though the structure is NCS. 

In recent years, many studies on the SOC terms in $\vb*{k}$-space Hamiltonians laid out a wider range of possibilities of spin-splitting patterns. For example, Samokhin derived the lowest-order allowed terms for all NCS point groups and showed that most of them differ from simple Rashba or Dresselhaus terms~\cite{Samokhin2009}. Vajna \emph{et al.} derived all terms up to third order in a subset of polar point groups that are relevant to metallic surfaces~\cite{Vajna2012}. Similarly, Cartoiaxo \emph{et al.} focused on a number of groups relevant to zinc blende quantum wells and derived the SOC Hamiltonians throughout the Brillouin zone~\cite{Cartoixa2006}. 
In Ref.~\cite{Acosta2021}, Acosta et al. took little group symmetries into account as well, thus showing different spin patterns can arise around different high symmetry points away from $\Gamma$, and in Ref.~\cite{Tao2024}, Tao approached the same problem in nonsymmorphic space groups.
There is also an intense interest in chiral materials, which lack any mirror planes in addition to $\mathcal{I}$, and their unique $\vb*{k}$-space spin textures' experimental signatures~\cite{Yazyev2023, Furukawa2021, Tan2022, Slawinska2023}. 
While all of these SOC effects emerge from the Russell-Saunders coupling  $\vb* L\cdot \vb*S$ between orbital and spin angular momenta, what terms are allowed in the solid's periodic Hamiltonian depends on its crystallographic symmetries, rendering the understanding of SOC a problem about the symmetry group of the material. 

In addition to fundamental scientific interest, SOC effects in electronic materials are also proposed as the basis for several spintronics applications~\cite{Zutic2004}. Two promising avenues for devices are the Datta-Das spin transistor~\cite{Datta1990,Datta2018}, which exploits the electric field manipulation of SOC-induced spin-splitting in semiconductors; and the Spin Hall Effect, which allows the generation of a pure spin current from the charge current in nonmagnetic materials~\cite{Manchon2015}. Additionally, tuning the relative strengths of Rashba and Dresselhaus effects (by strain, electric field, etc) makes it possible to obtain a `persistent spin helix' that circumvents the Dyakonov-Perel spin relaxation and leads to enhanced spin-current lifetime~\cite{Bernevig2006, Koralek2009, Tao2018, Tao2021, Lu2023}, paving the way to many spintronics devices. Finally, heterostructures of $s$-wave superconductors and semiconductors with Rashba SOC are a prime platform to realize Majorana modes with potential applications to quantum computation \cite{Lutchyn2010,Oreg2010}.

Together, these studies underline the need to systematically classify SOC terms in different NCS crystals in a comprehensive way. 
This is particularly important to enable the development of minimal models that can then be used to investigate superconductivity, charge density waves, or other related phenomena that are impacted by SOC. 
Achieving this goal requires an alternative approach to tabulating SOC terms by the NCS point groups in which they emerge. For example, instead of directly working with the 21 NCS crystallographic point groups themselves, one can focus on different types of symmetry breaking from a reference centrosymmetric group to NCS subgroups, and associate different SOC terms with specific broken symmetries. An intuitive way to classify and describe different types of non-magnetic symmetry breaking is to use electric (E) and electrotoroidal (ET) multipoles~\cite{Jackson2021, Dubovik1990}. Along these lines, Bhowal \emph{et al.} showed that nonmagnetic ferroelectrics host `magnetoelectric multipoles' with observable signatures in magnetic Compton scattering in reciprocal space~\cite{Bhowal2022}. Hayami \emph{et al.} provided a comprehensive classification of multipolar orders, their irreducible representations (irreps) of selected point groups, and $\vb*{k}$-space Hamiltonians they give rise to~\cite{Hayami2018}. While not referring to multipoles explicitly, Samokhin identified $\mathcal{I}$-odd irreps and SOC terms induced by their condensation using double group representations~\cite{Samokhin2019Pseudospin}. Despite the close connection between multipoles and point group representations, using irreps in conjunction with multipoles is in general more accurate, because multipole moments that are distinct in free space can mix when the rotational symmetry is lowered in the solid. 

Using $\mathcal{I}$-odd irreps of centrosymmetric crystals instead of the complete list of NCS point groups may, however, seem to have a shortcoming. Starting from a high-symmetry reference structure, each symmetry breaking order parameter, represented by a single irrep, gives rise to a single point (or space) group; but the opposite is not true. If a primary order parameter breaks all the symmetries that a secondary order parameter would break, then it would induce a nonzero magnitude of this secondary order parameter, too~\cite{Toledano1987}. This can lead to inconsistencies between the irrep and point group approaches if the secondary order parameters are not taken into account. Secondary order parameters also impact electronic structure features such as point nodes and nodal lines where the SOC-induced spin-splittings are zero. Thus, a complete description of SOC-induced spin-splitting in NCS requires inclusion of secondary order parameters as well.

In this study, we provide a unifying understanding of the problem of SOC in NCS crystals with time-reversal symmetry by combining point group irreps and multipolar representations of high-symmetry reference phases with $\vb*{k}$-space and minimal tight-binding (TB) models. 
Since every crystallographic point group in 3 dimensions is a subgroup of either the cubic group $m\bar{3}m$ ($O_h$ in Schoenflies notation) or the hexagonal group $6/mmm$ ($D_{6h}$ in Schoenflies notation), we choose these two groups as reference phases and consider their $\mathcal{I}$-odd irreps along with the corresponding multipoles. By constructing irrep projection operators~\cite{Dresselhaus2007}, we obtain leading order terms in reciprocal space SOC Hamiltonians corresponding to every $\mathcal{I}$-odd irrep of these two groups, and show that combinations of four distinct terms (Rashba, Dresselhaus, Weyl, and Ising) are sufficient to describe SOC to linear order in $\vb*{k}$. 
In addition to obtaining $\vb*{k}$-space Hamiltonians arising from  $\mathcal{I}$-odd irreps, we also use the projection operators to obtain minimal TB Hamiltonians associated with different irreps of $m\bar{3}m$.  
The advantage of this approach is that it enables associating SOC-induced terms in the Hamiltonian with specific OPs, and hence enables prediction of SOC Hamiltonians associated with any $\mathcal{I}$-odd OPs in both reciprocal-space band structures and TB Hamiltonians, developing a minimal yet complete classification scheme for SOC effects. 

Additionally, in order to bridge the gap between irrep-based and point group-based approaches, we tabulate all secondary $\mathcal{I}$-breaking irreps that are condensed when the primary order parameter transforms as a multi-dimensional irrep. 
This enables identifying the spin-splitting nodal features (points, lines or surfaces along which the SOC-induced splitting of bands with opposite spins vanishes) near the zone center from the reciprocal space Hamiltonians, as well as the nodal features  from the TB Hamiltonians. We thus provide a complete list of symmetry-imposed nodes and nodal lines near the zone center in NCS crystals. 
Additionally, we distinguish between three different types of symmetry-imposed nodal lines. The first type of nodal lines originate from 2-dimensional fermionic irreps along high-symmetry directions in the BZ, and thus is imposed by crystallographic symmetries even when $\mathcal{T}$ is absent. The second type of nodal lines arise between two complex-conjugate 1-dimensional irreps along high-order improper rotational axes (i.e., $\bar{4}$ or $\bar{6}$) that pass through a TRIM point, where the two irreps ``stick together'' along the entire axis as a result of the combination of improper rotational symmetry and $\mathcal{T}$. The third type of nodal lines occur in systems with a single mirror plane $m$, where the nodal lines originate from $\mathcal{T}$-imposed degeneracies between two complex-conjugate 1-dimensional irreps at the $\Gamma$ point and necessarily extend into finite-$\vb*{k}$ points on the mirror plane. The last type of nodal lines are not fixed along specific paths by crystallographic symmetries, but are determined by the electronic Hamiltonians which necessarily have real roots at finite-$\vb*{k}$ points on the mirror plane. We illustrate this type of nodal lines in a specific case, showing the transitions in topology under an external magnetic field perpendicular to the mirror plane.

We also show that an external magnetic field can be used to induce a topological phase transition in nonmagnetic NCS metals, in a way similar to altermagnets \cite{Fernandes2024}. 
Finally, to bridge our analysis with real materials applications, in the final part of this paper, we provide a list of previously synthesized examples of NCS materials for each point group that we obtain through a materials database search. 

Where our work overlaps with earlier studies, these results are in agreement with only a few exceptions. Additionally, our approach allows building connections that might be not as apparent in these earlier studies. For example, Ref.~\cite{Acosta2021} makes only limited use of irreps, but instead, relies more heavily on different point groups to classify different SOC terms. This leads to a disagreement between our results and their Fig.~3, where, for example, the Ising type SOC is not identified as a separate effect, and the Dresselhaus effect is not listed for certain point groups. Furthermore, the approach of studies such as Ref.~\cite{Acosta2021} does not distinguish between primary and secondary order parameters present in a low symmetry group, and which makes it challenging to capture certain subtleties about the nodal features in the band structure we discuss in Sec.~\ref{sec:secondary}.

This paper is organized as follows: in Section~\ref{sec:cubic}, we present a group theoretical analysis starting from the cubic parent group $m\bar{3}m$ and provide expressions for different spin-splitting terms as basis functions of $\mathcal{I}$-breaking irreps of this group. In Section~\ref{sec:hex}, we repeat the same process starting from the hexagonal parent group $6/mmm$. In Section~\ref{sec:multipoles}, we discuss the E/ET-multipoles in connection with irreps and spin-splitting. In Section~\ref{sec:tight binding}, we build TB models corresponding to each irrep in $m\bar{3}m$, and finally, in Section~\ref{sec:secondary}, we identify nodal features in NCS crystals' band structures, discuss the effect of secondary order parameters, and provide a list of materials examples. We conclude with a summary and conclusions in Section~\ref{sec:conc}.

\section{Spin-Splitting Hamiltonians in NCS Subgroups of the Cubic Parent Phase}
\label{sec:cubic}

\begin{figure*}
\includegraphics[width=0.8\linewidth]{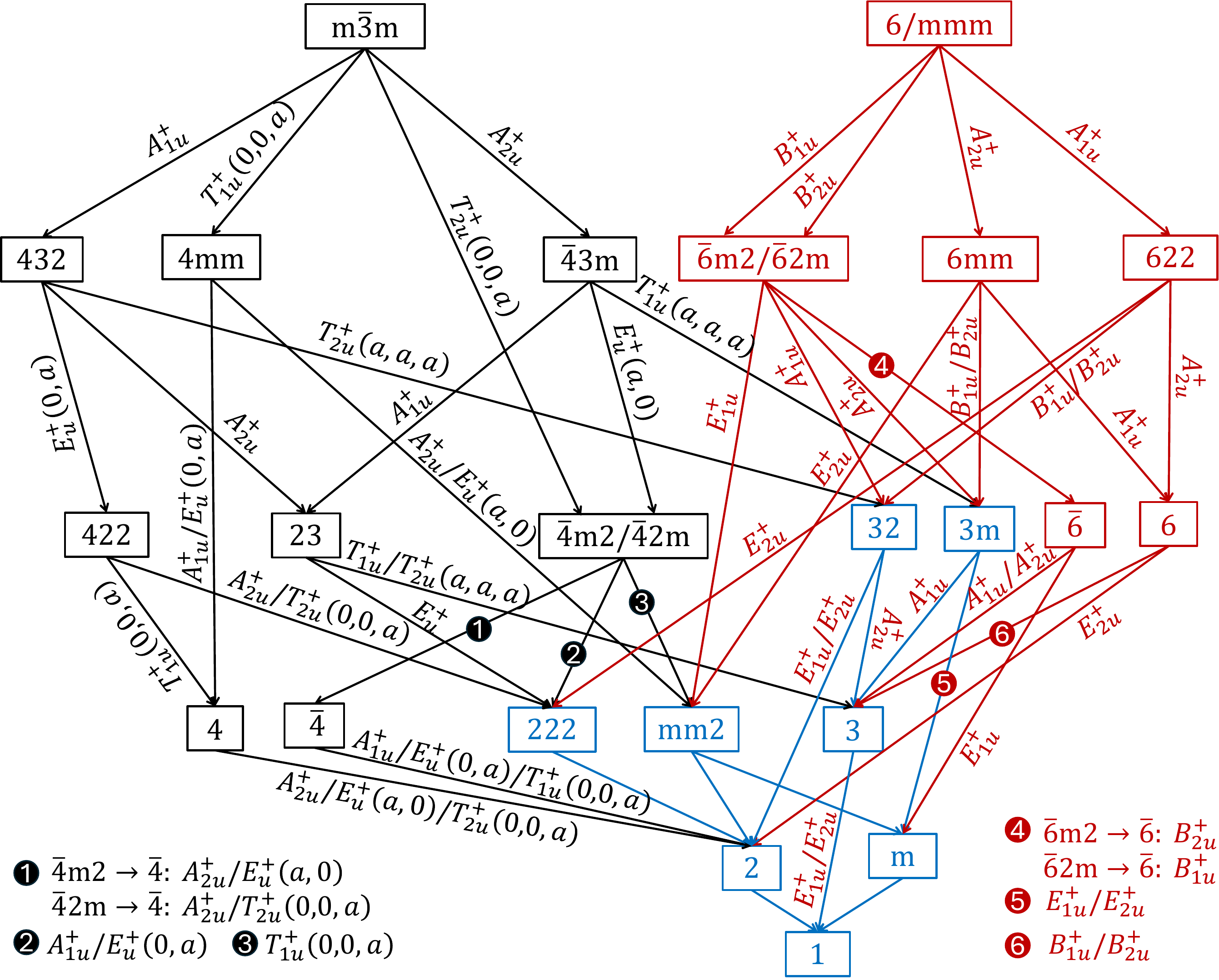}
\caption{Hierarchy tree of all NCS crystallographic point groups as subgroups of $m\bar{3}m$ or $6/mmm$. Each group is connected to its maximal subgroups by arrows, labeled with specific $\mathcal{I}$-odd irreps and their directions that break the corresponding symmetries. All irreps are that of the parent groups $m\bar{3}m$ (black) or $6/mmm$ (red). Irrep directions are not specified for connections between orthorhombic and lower groups, as multiple choices are possible. Common subgroups of $m\bar{3}m$ and $6/mmm$ are colored in blue.}
\label{fig:pg_tree}
\end{figure*}

The SOC Hamiltonian $\mathcal{H}_{SOC}=\lambda \vb*{L}\cdot\vb*{S}$ does not break either $\mathcal{T}$ or any crystallographic symmetries, since both the orbital ($\vb*{L}$) and the spin ($\vb*{S}$) angular momenta are $\mathcal{T}$-odd axial vectors, their dot product is a simple scalar. It breaks spin-only rotational symmetry, effectively imprinting the crystalline symmetries onto the spin degrees of freedom. Once projected onto the electronic bands, this term gives rise to effective terms in the reciprocal-space Hamiltonian that couple specific components of the wavevector $\vb*{k}$ with the electron spin represented by the Pauli spin-matrices $\vb*{\sigma} = [\sigma_x, \sigma_y, \sigma_z]$.

There are many past studies that list allowed SOC terms in different point groups (for example, Ref.~\cite{Samokhin2009}). Instead of associating these terms with different point groups, we choose to study how each different SOC term transforms under the symmetry operations of a high-symmetry centrosymmetric parent point group (e.g., cubic $m\bar{3}m$ or hexagonal $6/mmm$), and assign them $\mathcal{I}$-odd irreps of that group. In this and the next section, this approach will enable us to obtain a list of unique and distinct SOC terms, and avoid complications in low-symmetry groups where multiple such terms may coexist. Since every NCS subgroup of the parent group can be accessed by a combination of $\mathcal{I}$-odd irreps (Fig.~\ref{fig:pg_tree}), there are fewer distinct types of SOC terms in this classification scheme than the number of NCS point groups. For example, the 16 NCS subgroups of $m\bar{3}m$ can be reached by different combinations of 5 $\mathcal{I}$-odd irreps of it, as shown in Fig.~\ref{fig:pg_tree}, and hence can host only 5 distinct types of SOC terms. 

We start our analysis by elucidating the symmetry properties of the nine bilinear terms of the form $k_i \sigma_j$ ($i,j\in\{x,y,z\}$) that are linear in both the wavevector $\vb*{k}$ and the spin operator $\vb*{\sigma}$, and classifying them using the irreducible representations (irreps) of the point group. For this, we first consider the 9-dimensional representation $\Gamma^{k\sigma}$ according to which the nine $k_i \sigma_j$ terms transform. We first derive all the results for the cubic point group $m\bar{3}m$, and discuss the case of the other point groups that are subgroups of the hexagonal point group $6/mmm$ in Sec. \ref{sec:hex}.
In $m\bar{3}m$, the wavevector $\vb*{k}$, which is a $\mathcal{T}$-odd polar vector, transforms as the $T_{1u}^-$ irrep. Here, we follow the convention that the $\mp$ superscripts in point group irreps represent whether the irreps are $\mathcal{T}$-odd or even. The parity under $\mathcal{I}$ is represented in the subscript, where $g$ and $u$ represent $\mathcal{I}$-even (gerade) and odd (ungerade) irreps, respectively. The Pauli spin-matrices $\vb*{\sigma}$, represented as a classical $\mathcal{T}$-odd axial vector, transforms as $T_{1g}^-$ \footnote{This approach of representing spin as a classical vector allows us to use crystallographic groups for the rest of our analysis. It is also possible to use a spinor representation to represent the true spin-1/2 nature of the electron at this point, however, this requires moving onto the double groups, and in the end, gives identical results.}. The (reducible) representation $\Gamma^{k\sigma}$ is the direct product of $\Gamma^k=T_{1u}^-$ and $\Gamma^\sigma=T_{1g}^-$: 
\begin{equation}
\begin{aligned}    \Gamma^{k\sigma}&=\Gamma^{k}\otimes\Gamma^{\sigma}\\
        &=T^{-}_{1u} \otimes T^{-}_{1g}\\
        &= A^{+}_{1u} \oplus E^{+}_{u} \oplus T^{+}_{1u} \oplus T^{+}_{2u}
\end{aligned}
\label{eqn:bilinearirrep}
\end{equation}
As expected, all four irreps in this expansion are $\mathcal{T}$-even but $\mathcal{I}$-odd, so $\mathcal{I}$ needs to be broken for $k_i \sigma_j$ bilinears to appear in the reciprocal-space Hamiltonian. 

In the remainder of this section, we derive low-energy $\vb*k$-space Hamiltonian terms that transform as each $\mathcal{I}$-odd irrep of the cubic point group $m\bar{3}m$ ($O_h$) for a one-band model. A common example of an $\mathcal{I}$-breaking order in solids is a ferroelectric polarization, or in general, any structural order parameter (OP) that transforms as a polar vector. The polar OP $\vb*{P}$ transforms as the $T^{+}_{1u}$ irrep, and gives rise to the Rashba (R) spin-orbit coupling terms in the Hamiltonian proportional to $\vb*{P}\cdot\left(\vb*{k}\times \vb*{\sigma}\right)$. The fact that $T^{+}_{1u}$ appears in Eq.~\ref{eqn:bilinearirrep} means that the Hamiltonian of the parent cubic group $m\bar{3}m$ has terms that are of the form $\mathcal{H}_{R}(\vb*{P}) = \vb*{P} \cdot \vb*{h}_{R}$, where $\vb*{h}_{R}$ is a 3-component $\mathcal{T}$-even vector itself that is linear in momentum and linear in spin, 
$$\vb*{h}_{R}\equiv \alpha_R\begin{bmatrix}
     k_y\sigma_z -k_z\sigma_y\\
     k_z\sigma_x -k_x\sigma_z\\
     k_x\sigma_y -k_y\sigma_x
\end{bmatrix}.$$
Below a phase transition where a polar OP $\vb*{P}$ becomes nonzero, a Rashba effect linearly proportional to the OP emerges, and its explicit form depends on the direction of $\vb*{P}$. Equivalently, the three components of the vector $\vb*{h}_{R}$ transform as the $T_{1u}^+$ irrep, or as a polar vector. What this means is that the three components of $\vb*{h}_{R}$  switch places and signs according to the matrices of the $T_{1u}^+$ irrep under the point group operations. While these matrices are provided in standard group theory texts \cite{Aroyo2006}, we include a character table in Appendix~\ref{sm:character} for completeness. This result can be obtained by trial and error and considering how the wavevector and spin rotate under point group operations, and we show more details in Appendix~\ref{sm:rashba}. A more formal way to derive such terms for any irrep is using the projection operators, as described below. 

\begin{figure*}
\includegraphics[width=\linewidth]{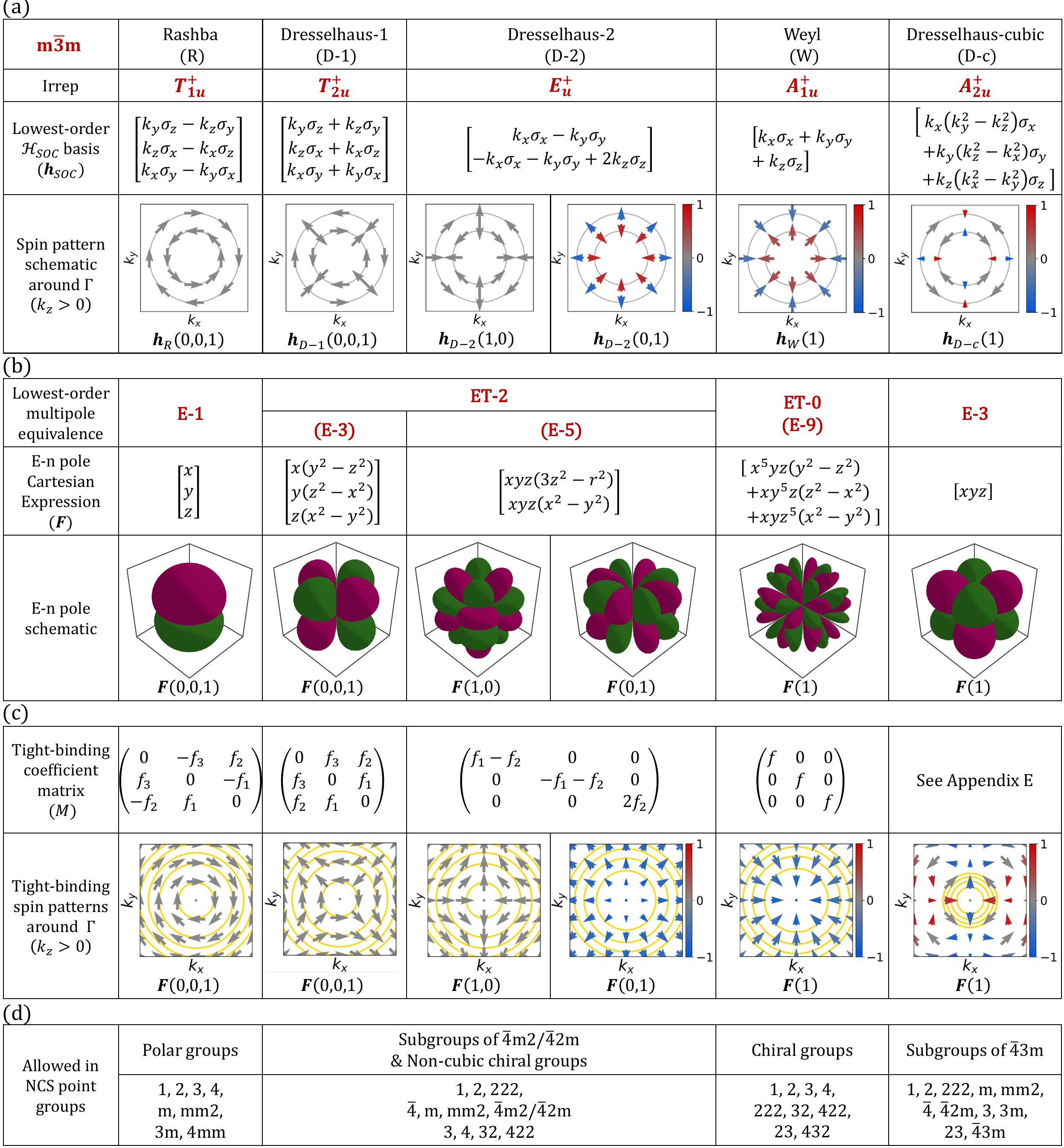}
\caption{Classification of different $\mathcal{H}_{SOC}(\vb*{k},\vb*{\sigma})$ corresponding to the $\mathcal{I}$-odd irreps of the most symmetric cubic group, $m\bar{3}m$. The Hamiltonian is given by this term multipled by a (possibly multi-component) order parameter that transforms as the corresponding irrep. (a) Irreps of $m\bar{3}m$ according to which each class of $\mathcal{H}_{SOC}$ terms transform, and the lowest-order bases listed in pseudo-vector form. Each set of bases are chosen to be orthogonal and to span the space of the particular irrep. As a result, the dimensionality of the vector of bilinears is determined by the dimensionality of the irrep. Schematics of resultant spin patterns are plotted on $k_x$-$k_y$ plane around $\Gamma$ with $k_z$ slightly above 0 in order to demonstrate the feature of out-of-plane spin components. The direction of the arrows corresponds to the in-plane components, whereas their color denotes the out-of-plane components (red for spin-up; blue for spin-down; gray denotes no out-of-plane component). (b) Lowest-order isomorphic E/ET-multipole equivalence for each class of $\mathcal{H}_{SOC}$ that transforms in the same way in free space as the lowest-order $\mathcal{H}_{SOC}$ bases. For ET-multipoles, the lowest-order E-$n$ pole equivalence in $m\bar{3}m$ is listed in parentheses. Explicit mathematical expressions (normalization factor omitted) and schematics are listed for each E-multipole. (c) Coefficient matrices $M$ of TB models for each class (see Section~\ref{sec:tight binding} for details) and resultant spin patterns around $\Gamma$ in the lowest spin-polarized TB band. Color scheme represents out-of-spin components; golden circles represent equi-energy contours with a fixed interval. (d) NCS point groups that are subgroups of $m\bar{3}m$ in which each class of spin terms are allowed to emerge.}
\label{fig:cubic}
\end{figure*}

\begin{figure}
\includegraphics[width=1.0\linewidth]{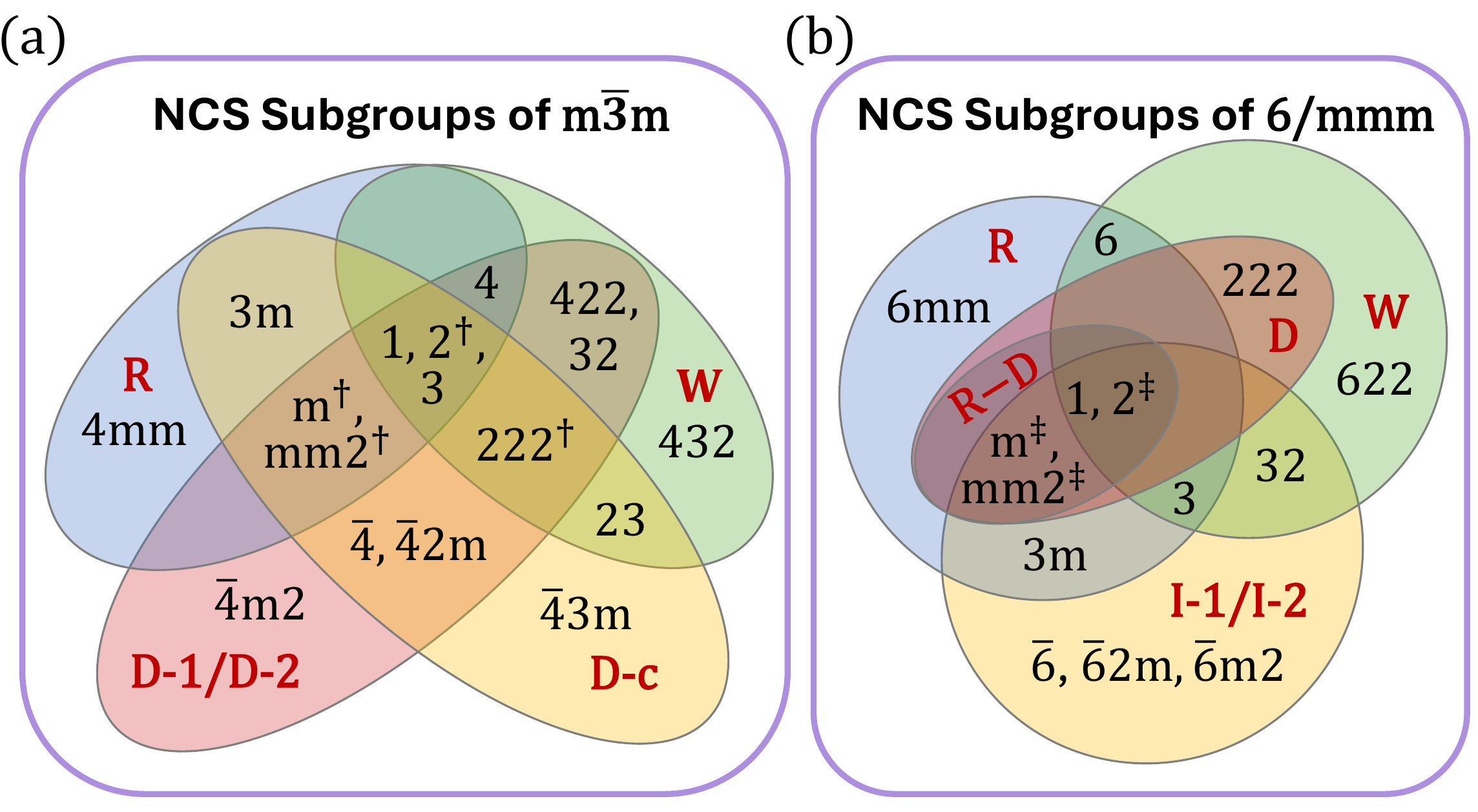}
\caption{(a) NCS subgroups of $m\bar{3}m$ and allowed spin-splitting terms therein. Coordinate axes are chosen in the same way as in Fig.~\ref{fig:cubic}. For monoclinic and orthorhombic groups (labeled with $\dagger$) that have multiple possible symmetry alignments, D-c is specifically allowed when $m\perp\langle 110\rangle$ and $2\parallel\langle 100\rangle$ in the coordinates of $m\bar{3}m$. (b) NCS subgroups of $6/mmm$ and allowed spin-splitting terms therein. Definitions of spin-splitting terms in the hexagonal case, including the Ising SOC, are given in Fig.~\ref{fig:hex}. For subgroups of $mm2$ (labeled with $\ddagger$), alignments of symmetries can be different for R-D or R or D terms to emerge.}
\label{fig:pg_venn}
\end{figure}

Besides the polarization $\vb*{P}$ , the cubic group has four other $\mathcal{I}$-odd irreps ($A_{1u}^+$, $A_{2u}^+$, $E_{u}^+$, and $T_{2u}^+$) that are non-polar and transform as different irreps than $T_{1u}^+$ . Each should lead to different combinations of bilinears $k_i\sigma_j$ (or higher order combinations) in the Hamiltonian. The number of linearly independent bilinears that transform to each other according to each irrep is determined by the dimensionality of the irrep. The form of these bilinears, which can be considered as sets of basis functions $\vb*{F}$ of the corresponding irreps, can be found using the irrep projection operators. We consider the projection operator $\hat{P}^{\Gamma_n}$ that projects the Hamiltonian it acts on onto the $\Gamma_n$ irrep. It is given (up to a multiplicative constant) by
\begin{equation}
\label{eqn:PO}
\hat{P}^{\Gamma_n}  = \sum_R \chi^{\Gamma_n}(R)\hat{P}_R ,
\end{equation}
where the sum is over all symmetry operations $R$ in the point group. $\hat{P}_R$ is the symmetry operator and $\chi^{\Gamma_n}(R)$ is the character of $R$ in the irrep $\Gamma_n$. This form of the irrep projection operators follows a standard definition~\cite{Dresselhaus2007}, but it is not commonly applied on band Hamiltonians. Acting with $\hat{P}^{\Gamma_n}$ on the sum of all bilinears $k_i \sigma_j$ gives only the combinations of bilinears that transform as $\Gamma_n$. We perform this analysis for all $\mathcal{I}$-odd irreps with the help of the INVARIANTS~\cite{INVARIANTS_1} tool from the ISOTROPY Software Suite, which adopts a different algorithm~\cite{INVARIANTS_2} that gives equivalent results to Eq.~\ref{eqn:PO}. The results are shown in Fig.~\ref{fig:cubic}(a).

As expected, the three components of the linear Rashba term (which we henceforth denote as R) form the basis of the 3D irrep $T^+_{1u}$, which transform the same way as the three components of a polar vector ($\vb*{F}=[x,y,z]$).
The reciprocal space spin pattern for a pair of parabolic subbands that would have been degenerate in the absence of SOC, shown in the first column of Fig.~\ref{fig:cubic}(a), is the typical Rashba pattern, where the spin polarization is always normal to the wavevector. In the spin-pattern figures, the gray arrows represent spins on the $xy$ plane, whereas blue and red colors represent an out-of-plane ($z$) component.

The bilinears that transform as $T_{2u}^+$ and $E_u^+$ correspond to two different types of linear Dresselhaus terms (shown in the second and third columns), which we denote as D-1 and D-2. We classify these irreps together under the same name Dresselhaus, not only because the spin patterns they give rise to are similar, but also because they are components of the same $\mathcal{I}$-odd $l=2$ irrep of the $O(3)$ group in free space. In all possible spin patterns resulting from both irreps, there are regions in each subband where the spin directions alter between pointing into and pointing out of the unit sphere in the $\vb*{k}$-space around $\Gamma$. In particular, the first component of $E_u^+$ ($\vb*{h}_{D-2}(1,0)$) gives rise to a pattern which is just 45$^\circ$-rotated from the one generated by the third component of $T_{2u}^+$ ($\vb*{h}_{D-1}(0,0,1)$). Meanwhile, $\vb*{h}_{D-1}(1,1,1)$ gives rise to a pattern which is identical in shape to the pattern generated by $\vb*{h}_{D-2}(0,1)$ but with the $z$-axis pointing along [111] (body diagonal) in the cubic coordinate. Hence, these two effects differ by the choice of the cartesian coordinate axes. An additional reason to group the D-1 and D-2 terms together will be discussed in Section~\ref{sec:multipoles}, where we show that the 5 total components of these terms together transform as the 5 components of the electric toroidal quadrupole in free space. 

The last type of bilinear spin-splitting Hamiltonian is the Weyl-like (W) spin-splitting term \cite{Acosta2021}. Unlike the cases discussed so far, the linear SOC Weyl effect is completely isotropic even though it is gyrotropic~\cite{Fu2015, Watanabe2018}, giving rise to a radial spin pattern with spin polarization collinear with $\vb*{k}$ at all points. The corresponding Hamiltonian can be expressed as the dot product of the polar and axial vectors $\vb*{k}$ and $\vb*{\sigma}$: $\mathcal{H}_{W}=\alpha_{W}\vb*{k}\cdot\vb*{\sigma}$. Here, the coefficient $\alpha_{W}$ is proportional to the OP that transforms as the irrep $A_{1u}^+$, which corresponds to a chiral order parameter. In tetragonal and lower chiral groups, this term always coexists and is mixed with $\vb*{h}_{D-2}(0,1)$, leading to two independent terms, $(k_x\sigma_x+k_y\sigma_y)$ and $k_z\sigma_z$~\cite{Yazyev2023}, similar to the Weyl term in hexagonal and trigonal systems which will be discussed in the next section.

In the cubic group $m\bar{3}m$, there are 5 $\mathcal{I}$-odd irreps in total, but only 4 of them appear in the representation $\Gamma^{k\sigma}$ of  $k_i\sigma_j$ bilinear SOC Hamiltonians (Eq.~\ref{eqn:bilinearirrep}), indicating that there is no combination of bilinear terms that transform as the $A^+_{2u}$ irrep. Hence, the simplest $A^+_{2u}$ spin-splitting term has to include a higher order of $\vb*k$. 
The next order of spin-splitting terms are quadratic in $\vb*{k}$, and hence break $\mathcal{T}$. While quadratic $k_ik_j\sigma_k$ terms are responsible for interesting spin-splitting patterns in altermagnets~\cite{Smejkal2022_1,Smejkal2022_2,Jungwirth2026}, they do not require SOC and break time-reversal symmetry $\mathcal{T}$. SOC terms that are cubic in $k_i$ and linear in $\sigma_j$, however, obey $\mathcal{T}$, and the irreps of all possible $k_i k_j k_k \sigma_l$ terms are given by the product decomposition
\begin{equation}
\begin{aligned}
\Gamma^{k^3\sigma} &=\Gamma_{s}^{k^3}\otimes\Gamma^{\sigma} \\ 
&=[T^-_{1u}]^3_{s} \otimes T^-_{1g} \\
&=2A^+_{1u} \oplus A^+_{2u} \oplus 3E^+_{u} \oplus 3T^+_{1u} \oplus 4T^+_{2u}. 
\end{aligned}
\label{eqn:kcubed}
\end{equation}
Here, the square brackets $[~]^n_s$ represent the symmetrized $n$-th power, which is needed because $k_i k_j = k_j k_i$~\cite{Hecker2024}.
$\Gamma^{k^3\sigma}$ is a 30-dimensional reducible representation, which is a sum of 13 irreps of $m\bar{3}m$, including every $\mathcal{I}$-odd irrep at least once. This means that any $\mathcal{I}$-breaking order leads to the emergence of at least one term that is cubic in $\vb*{k}$ and linear in $\vb*{\sigma}$. 
$A^+_{2u}$ appears in Eq.~\ref{eqn:kcubed} once, thus there is only one $k_i k_j k_k \sigma_l$ term that transforms as this irrep. This term, shown in the last column of Fig.~\ref{fig:cubic}, is present and well studied in the zinc blende structure with point group $\bar{4}3m$, where it is the lowest-order SOC term allowed. Following the convention of referring to this as a Dresselhaus type of term, but at the same time distinguishing it from the lower-order Dresselhaus terms discussed before, we call this term the `Dresselhaus-cubic' (D-c) term.

Since the remaining 12 irreps contributing to $\Gamma^{k^3\sigma}$ give rise to cubic-in-$\vb*{k}$ $\mathcal{H}_{SOC}$ terms belonging to the previously listed classes, we do not discuss them further. We provide a complete list of these terms in Appendix~\ref{sm:hamiltonian_3rd}. 
All of these results are consistent with the earlier work of Watanabe and Yanase~\cite{Watanabe2018} where they overlap. 
In general, more than one SOC term may coexist in a given noncentrosymmetric point group. We summarize the allowed SOC terms in all NCS point groups that are subgroups of $m\bar{3}m$ in a Venn diagram in Fig.~\ref{fig:pg_venn}(a). The connection between these subgroups and the $\mathcal{I}$-odd irreps of the cubic group is represented by the hierarchy tree shown in Fig.~\ref{fig:pg_tree}.

We note that the analyses in this paper do not require an explicit discussion of the double groups and their representations~\cite{Bradley2011} (except for a brief discussion in Section~\ref{sec:secondary}). Even though electrons (and hence the fermionic annihilation and creation operators we use in Section~\ref{sec:tight binding}) attain a phase under a 360$^\circ$ rotation, all the terms in the SOC Hamiltonians are bilinears of fermionic operators and hence do not attain a phase under 360$^\circ$ rotations. This enables them to be classified using single-valued point groups and their representations.

\section{Spin-Splitting Hamiltonians in NCS Subgroups of the Hexagonal Parent Phase}
\label{sec:hex}

Not all NCS groups are subgroups of the high-symmetry cubic group $m\bar{3}m$. In this section, we analyze the NCS subgroups of the high-symmetry hexagonal group $6/mmm$. As shown in Fig.~\ref{fig:pg_venn}, the combination of subgroups of the cubic and hexagonal parent groups we consider covers all NCS crystallographic point groups, and thus contain all types of SOC allowed. On the right half of Fig.~\ref{fig:pg_tree}, we show the transition paths connecting the NCS subgroups of $6/mmm$ to its  $\mathcal{I}$-odd irreps.

\begin{figure*}
\includegraphics[width=\linewidth]{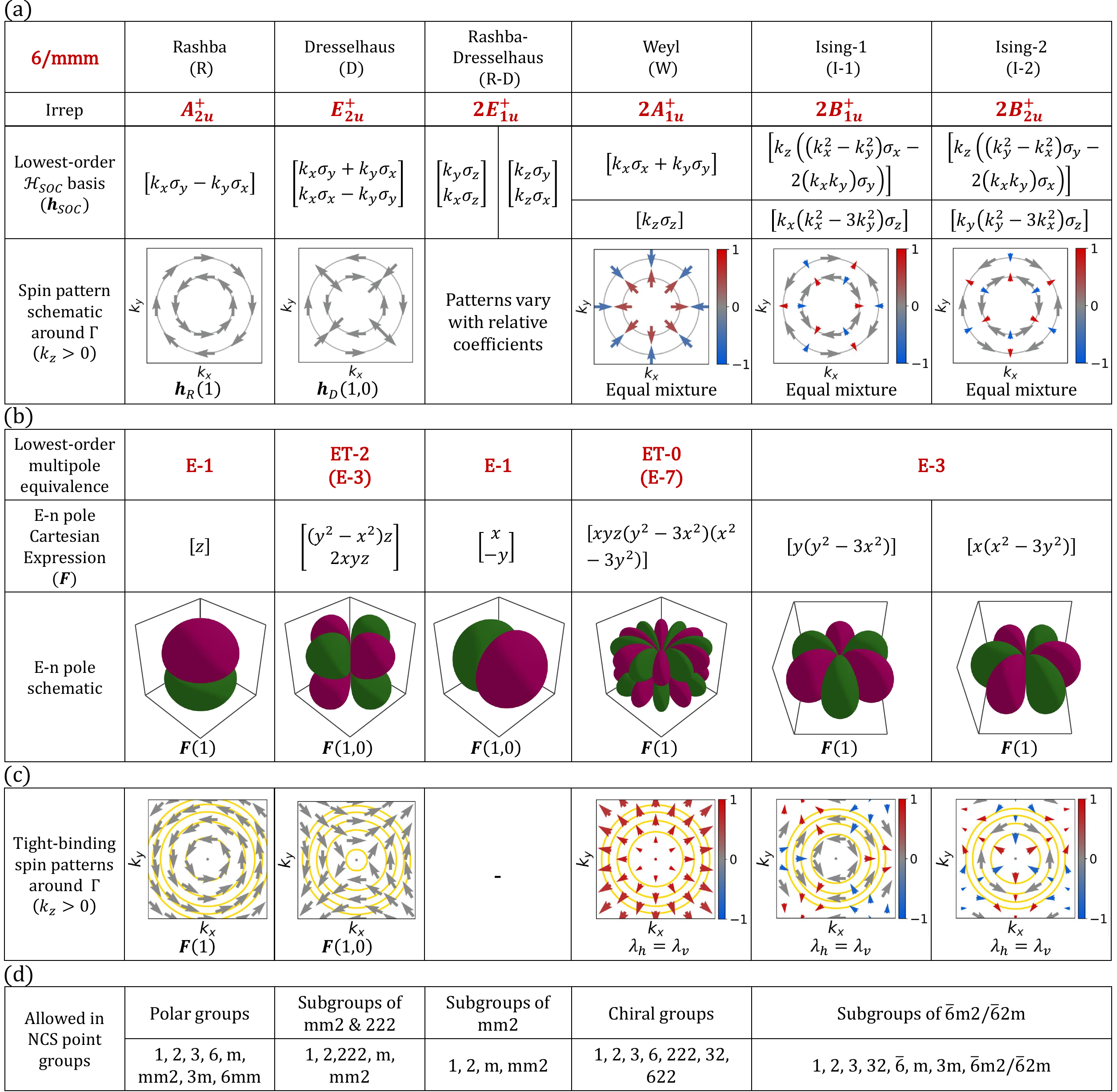}
\caption{Classification of the different $\mathcal{H}_{SOC}(\vb*{k},\vb*{\sigma})$ terms in the form of vectors corresponding to the $\mathcal{I}$-odd irreps of the most symmetric hexagonal point group, $6/mmm$. The Hamiltonian is given by the vector multiplied by a (possibly multi-component) order parameter that transforms as the corresponding irrep. (a) Irreps of $6/mmm$ according to which each class of $\mathcal{H}_{SOC}$ terms transform. Bases are chosen to be orthogonal and to span the space of the corresponding irrep, with schematics of the resultant spin patterns shown on $k_x-k_y$ plane around $\Gamma$. $k_z$ is set to be slightly above 0 in order to demonstrate the out-of-plane spin features (red for spin-up; blue for spin-down). (b) Lowest-order isomorphic E/ET-multipole equivalence for each class of $\mathcal{H}_{SOC}$ which transforms in the same way in free space as the lowest-order $\mathcal{H}_{SOC}$ bases. For ET-multipoles, the lowest-order E-n pole equivalence in $6/mmm$ is listed in parentheses. (c) Corresponding spin patterns around $\Gamma$ in the lowest spin-polarized band for each class from TB models (see Appendix~\ref{sm:tight binding_hex}). Color scheme represents out-of-spin components; golden circles represent equi-energy contours with a fixed interval. (d) NCS point groups that are subgroups of $6/mmm$ in which each class of spin terms are allowed to emerge.}
\label{fig:hex}
\end{figure*}

To obtain the complete set of SOC terms represented by the $\mathcal{I}$-odd irreps of $6/mmm$, we follow the same procedure as in the cubic case. The wavevector $\vb*{k}$ transforms as $E_{1u}^-\oplus A_{2u}^-$ ($\{x,y\}\oplus\{z\}$) in $6/mmm$, while the Pauli matrices $\vb*{\sigma}$ transform as $E_{1g}^-\oplus A_{2g}^-$ ($\{\sigma_x,\sigma_y\}\oplus\{\sigma_z\}$). Following the same line of analysis in Section~\ref{sec:cubic}, we find the reducible representation $\Gamma^{k\sigma}$ of the linear-in-$\vb*{k}$ terms in $6/mmm$ as: 
\begin{equation}
\begin{aligned}    \Gamma^{k\sigma}&=\Gamma^{k}\otimes\Gamma^{\sigma}\\
        &= (E_{1u}^-\oplus A_{2u}^-) \otimes (E_{1g}^-\oplus A_{2g}^-)\\
        &= 2A^{+}_{1u} \oplus A^{+}_{2u} \oplus 2E^{+}_{1u} \oplus E^{+}_{2u},
\end{aligned}
\label{eqn:bilinearirrep_hex}
\end{equation}
which is a 9-dimensional representation that breaks into 6 irreps of 4 distinct kinds. The classification of the $k_i\sigma_j$ bilinears that transform according to these irreps is shown in the first four columns in Fig.~\ref{fig:hex}(a). We name each class of $k_i\sigma_j$ terms in analogy to the cubic $m\bar{3}m$ system, as these terms can be understood as combinations of the Rashba, Dresselhaus, and Weyl SOC terms obtained in the $m\bar{3}m$ case. However, in the $6/mmm$ case, there is a clear difference between the terms with $\vb*{k}$ located on the $k_x$-$k_y$ plane and along the $k_z$ axis. This is a consequence of the characteristic $6$ or $\bar{6}$ axes present in hexagonal point groups, instead of the multiple diagonal $3$-axes in $m\bar{3}m$ that mixes $x,y,z$ in a cyclic way with each other.

Similarly to Eq.~\ref{eqn:bilinearirrep}, we find that there are $\mathcal{I}$-odd, $\mathcal{T}$-even irreps of $6/mmm$ missing in the decomposition of bilinear in Eq.~\ref{eqn:bilinearirrep_hex}, namely, the two one-dimensional irreps, $B^+_{1u}$ and $B^+_{2u}$. Moving on to the analysis of cubic-in-$\vb*{k}$ terms, we see two copies of each irrep appearing in the reducible representation $\Gamma^{k^3\sigma}$:
\begin{equation}
\begin{aligned}
\Gamma^{k^3\sigma}=&\Gamma_{s}^{k^3}\otimes\Gamma^{\sigma} \\ 
=&[E_{1u}^-\oplus A_{2u}^-]^3_{s} \otimes (E_{1g}^-\oplus A_{2g}^-) \\
=&\Bigl([E_{1u}^-]^3_s \oplus ([E_{1u}^-]^2_s\otimes A_{2u}^-) \oplus (E_{1u}^-\otimes[A_{2u}^-]^2_s) \\
&\oplus [A_{2u}^-]^3_s \Big)\otimes (E_{1g}^-\oplus A_{2g}^-) \\
=&4A_{1u}^+ \oplus 2A_{2u}^+ \oplus 2B_{1u}^+ \oplus 2B_{2u}^+ \oplus 5E_{1u}^+ \oplus 5E_{2u}^+ .
\end{aligned}
\label{eqn:kcubed_hex}
\end{equation}
Thus, the 30-dimensional representation  $\Gamma^{k^3\sigma}$  breaks into 20 irreps of $6/mmm$, 10 of which being one-dimensional and the other 10 being two-dimensional. 

The two copies of $B_{1u}^+$ ($B_{2u}^+$) correspond to two separate bases of $k_ik_jk_k\sigma_l$ terms that transform independently and identically as the irrep $B_{1u}^+$ ($B_{2u}^+$). Their explicit expressions are listed in the last two columns of Fig.~\ref{fig:hex}(a), along with the resultant spin pattern schematics. We refer to these terms as Ising (I) SOC. Ising SOC is discussed often in monolayer transition metal dichalcogenides, corresponding to an alternating out-of-plane magnetic field at the $K$ and $K'=-K$ points of the $k_z=0$ Brillouin Zone \cite{Xi2016, Zhou2016}. It is associated with the breaking of the in-plane mirror symmetries (i.e. mirror planes with in-plane normals) \cite{Zhang2023}, and not with the breaking of out-of-plane mirror or any rotational symmetries. This description is consistent with the $B_{1u}^+$ and $B_{2u}^+$ irreps of $6/mmm$, which preserve the $\bar{6}$ improper rotation (roto-inversion) and the out-of-plane mirror $/m$, but break half of the in-plane mirrors each. These two terms are the only terms allowed in systems with the improper rotation axis $\bar{6}$, similar to the Dresselhaus terms (D-1, D-2, D-c) in the cubic case, which are the only terms allowed in systems with the improper rotational axis $\bar{4}$.

\section{Multipole Representations of the Spin-Splitting Hamiltonians}
\label{sec:multipoles}

\begin{figure}
\includegraphics[width=0.8\linewidth]{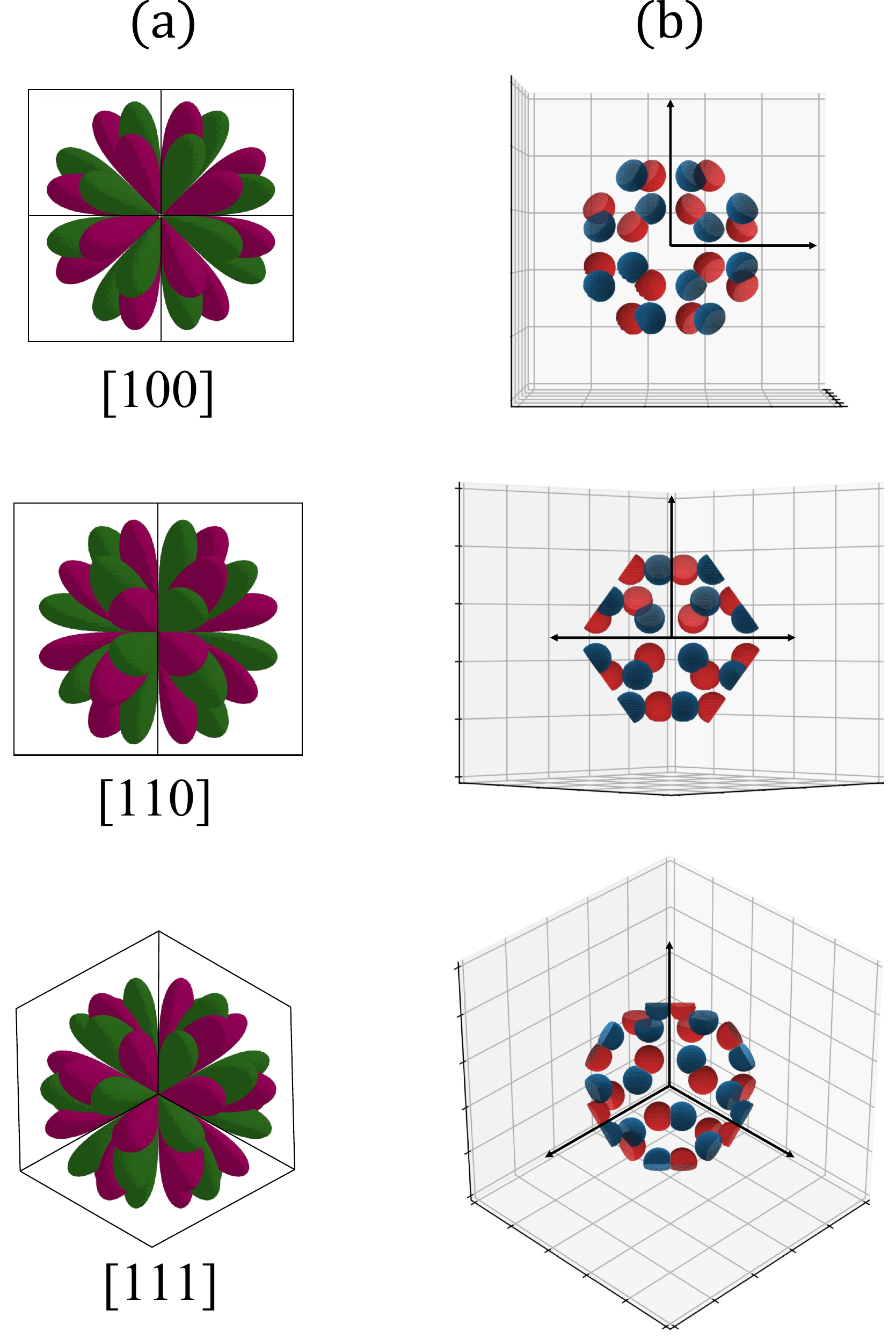}
\caption{(a) Schematics of the E-512 multipole (i.e., the $l=9$ electric multipole) corresponding to the Weyl SOC term of a cubic crystal, viewed from different crystal directions. Panel (b) shows the $A_{1u}^+$-projected charge density distribution around the central atom (Ce) in the hypothetical material LaCeBe$_{26}$ (chiral space group $F432$) calculated by DFT. Purple/red lobes represent charge-excessive regions and green/blue ones represent charge-deficient regions. Charge density is processed using the package ProDenCeR~\cite{Prodencer_1,Prodencer_2}.}
\label{fig:E512}
\end{figure}

Electric (E) and electrotoroidal (ET) multipoles, characterized by angular momentum labels $l$ and $m$, form the irreps of the point group of free space, $O(3)$~\cite{Jackson2021, Dresselhaus2007, Hayami2018}. In other words, they provide a complete and orthonormal basis for representing the angular dependence of $\mathcal{T}$-invariant order parameters and quantities like charge density. In a crystalline system, the reduction of the continuous rotational symmetry to a discrete one leads to a finite-order point group with a finite number of irreps. In this case, each irrep corresponds to an infinite number of multipoles. However, since the higher-order multipoles are often increasingly irrelevant because of the increasing number of nodes they have, a common practice is to associate each point group irrep with the lowest-order multipole that transforms as that irrep~\cite{Watanabe2018,Hayami2024}. This enables developing intuition about the physical meanings of different irreps of the crystallographic point groups and connect them with other experimental observables. Multipoles are often used to represent order parameters as well~\cite{Ederer2007, Schaufelberger2023}. 

A complete list of low-order multipoles' irreps and their basis functions for m$\bar{3}$m is tabulated in, for example, Ref.~\cite{Hayami2018}. In Fig.~\ref{fig:cubic}(b), we use this result to list the multipoles that are isomorphic to the different classes of SOC Hamiltonians corresponding to the $\mathcal{I}$-odd irreps of the cubic reference group $m\bar{3}m$.
Not surprisingly, the $\mathcal{I}$-odd multipoles appearing in Fig.~\ref{fig:cubic}(b) are either odd-ranked E-multipoles or even-ranked ET-multipoles, where the rank refers to the angular momentum label $l$ of the multipole. Aside from the Rashba (R) term, which is represented by the 3-component electric dipole (E-1), the two Dresselhaus terms D-1 and D-2 transform as the 5-component electric toroidal quadrupole (ET-2), which is split into two irreps (3-fold + 2-fold) due to the cubic crystal field. The cubic-Dresselhaus (D-c) term transforms as a specific component of the electric octupole (E-3), where the other 6 components of the octupole transform as irreps that overlap with the R and D-1 terms. 

Finally, the Weyl (W) term transforms as the electric toroidal monopole (ET-0), which is invariant under all rotations even though it is odd under $\mathcal{I}$. Recent studies explain ET-0 as a quantitative measure of chirality~\cite{Inda2024,Huang2026}. This implies that the Weyl-like spin pattern in reciprocal space is also a simple expression of chirality. 

When an OP represented by an E/ET-multipole condenses in a phase transition, it gives rise to the corresponding spin-splitting pattern in reciprocal space. For E-multipoles, visualization of $\mathcal{I}$-breaking OPs in real space is rather straightforward, as the OP is directly manifested in the charge-density distribution (Fig.~\ref{fig:cubic}(b)). 
ET-multipoles, on the other hand, are another unique set of bases that cannot be represented by the combination of E-multipoles (and thus the charge density distributions) \textit{in free space}, because ET-multipoles behave the same way as E-multipoles of the same order under all rotational symmetries, but have opposite parities under $\mathcal{I}$~\cite{Winkler2023,Winkler2025}.
This can be understood from the fact that there is no odd-order polynomial of cartesian coordinates that is $\mathcal{I}$-even, e.g., one cannot have a charge density distribution that has a dipolar form but is $\mathcal{I}$-even, as should be the case for ET-1. 
While this renders simple visualization of ET multipoles in free space impossible, the finite number of irreps of crystallographic point groups provides a loophole. Due to the finite order of crystallographic point groups, there is always a higher-order E-multipole that transforms as the same irrep as a specific ET-multipole. These E-multipoles can then be used to visualize and gain insight about the nature of the symmetry breaking pattern associated with an ET-multipole. 

We identified polynomials of $(x,y,z)$ that transform as a given irrep using the projection operator (Eq.~\ref{eqn:PO}) acting on different polynomials of increasing order. In Fig.~\ref{fig:cubic}(b), we list the lowest-order E-multipoles that transform as the same irrep as the ET-multipoles in parentheses, and show sketches of their isosurfaces. 
For example, the D-1 and D-2 terms induced by the electrotoroidal quadrupoles (ET-2) correspond to electric octupole (E-3) and dotriacontapole (32-pole, E-5) moments. 

The lowest-order E-multipole that transforms the same way as the electric toroidal monopole (ET-0) is the electric \textit{pentahectadodecapole} (512-pole, E-9). To visualize it, we use density functional theory (DFT) to obtain the charge density of a hypothetical system LaCeBe$_{26}$~\cite{lacebe26} in the chiral space group $F432$. We show in Fig.~\ref{fig:E512}, from different angles of view, the DFT charge density projected onto the $A_{1u}^+$ irrep~\cite{Prodencer_1, Prodencer_2}\footnote{The DFT calculation is conducted using the same settings documented in Appendix~\ref{sm:dft}, except for a cut-off energy of 380~eV, with on-site inter-electron repulsion of U=3~eV (7~eV) for La-$d$ (Ce-$f$) orbitals. Crystal structure is obtained from Materials Project Database (mp-1222939)~\cite{lacebe26} and fully relaxed.}. The reason of such a high-order E-multipole being necessary to represent ET-0 is that it must preserve all rotation symmetries while breaking $\mathcal{I}$ and all mirror symmetries. In $m\bar{3}m$, there are 24 improper rotations (including mirrors) to be broken, while there are 24 proper rotations to be preserved. Only a cubic harmonic with 48 lobes ($l$=9) can achieve both at the same time. 

A nonzero expectation value of ET-0 ($A_{1u}^+$) reduces the point group symmetry from $m\bar{3}m$ ($O_h$) to $432$ ($O$). Materials with point group $432$ would display a spin-splitting pattern that is completely isotropic in the lowest order of $\vb*{k}$. We performed a materials database search to find such a material, but found that this is an exceedingly rare point group in symmorphic crystals, and we could not identify any synthesized inorganic crystals with space group $P432$, $I432$, or $F432$. On the other hand, there are many experimentally observed compounds with nonsymmorphic space groups with $432$ point symmetry. In Appendix~\ref{sm:dft}, we report the electronic structure of Mo$_3$Al$_2$C with space group $P4_132$~\cite{Zhigadlo2018, Wu2024, Wu2025} from DFT using the Vienna Ab initio Simulation Package (VASP)~\cite{Kohn1965,Kresse1996_DFT1,Kresse1996_DFT2,Blochl1994PAW1,kresse1999PAW2,Perdew1996_PBE,Perdew2008_PBEsol}, and show that it exhibits Weyl-type spin-splitting pattern without any Rashba or Dresselhaus effects present. 

A similar multipolar analysis can be performed for the $\mathcal{I}$-odd irreps of the hexagonal group $6/mmm$. The results are shown in Fig.~\ref{fig:hex}(b). Note that the two Ising SOC terms correspond to components of the E-3 multipole (i.e., an electric octupole) that do not appear as leading-order multipoles in any of the SOC terms in the NCS subgroups of the cubic group $m\bar{3}m$.

\section{Tight-Binding Models for Spin-Splitting Multipoles}
\label{sec:tight binding}

\begin{figure}
\includegraphics[width=0.95\linewidth]{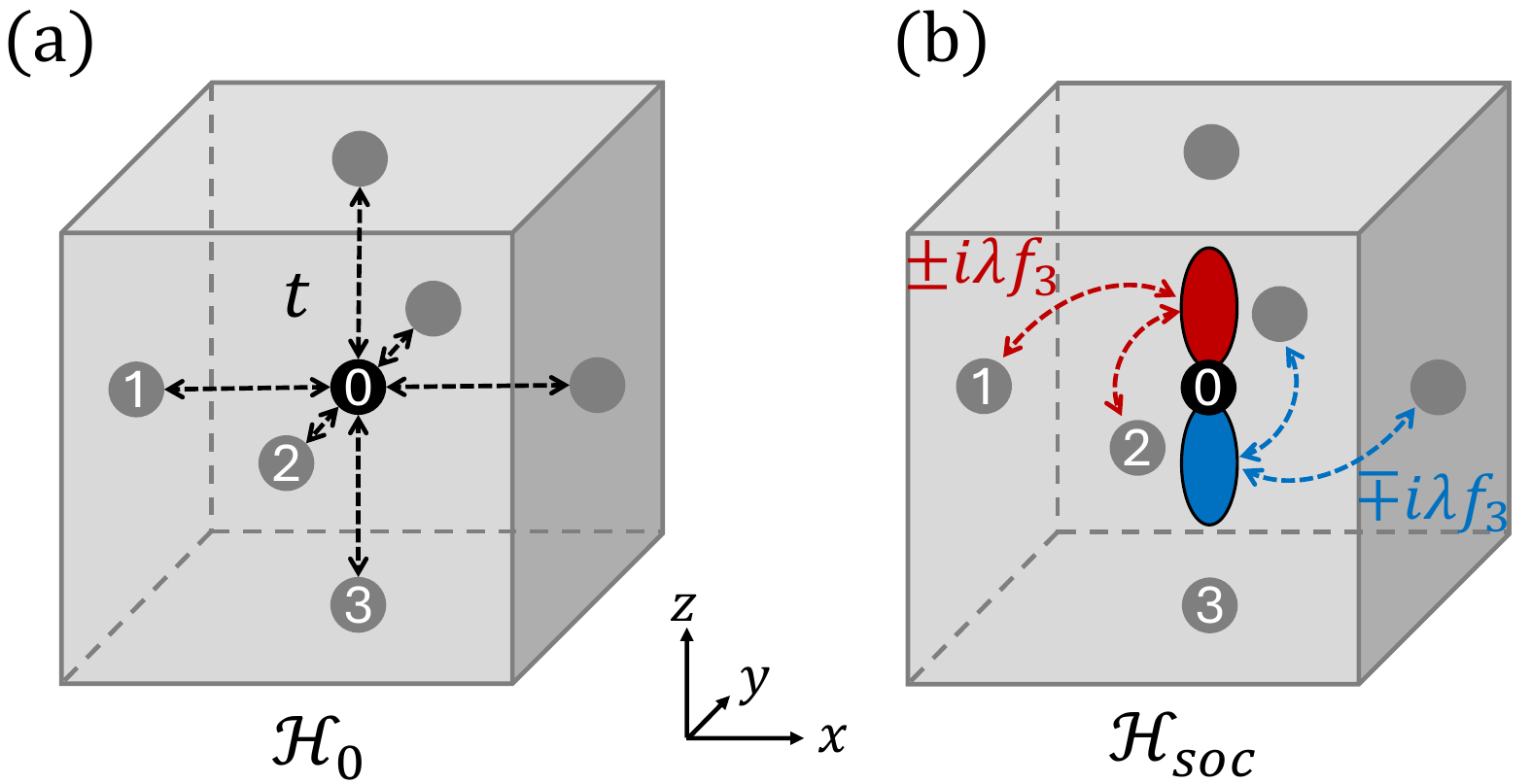}
\caption{Example of a TB model constructed in the parent phase $m\bar{3}m$ corresponding to the SOC term associated with the electric dipole $T_{1u}^+(0,0,1)$. Hoppings (dashed arrows) are considered between site 0 and its nearest neighbors (sites 1, 2, 3). (a) The spinless part $\mathcal{H}_{0}$ respects full symmetries of $m\bar{3}m$ and (b) the SOC part $\mathcal{H}_{SOC}$ reproduces the symmetries of the electric multipole present in the system. The same unit cell is used to construct TB models for all but the Dresselhaus-cubic (D-c) SOC terms (See Appendix~\ref{sm:D_c}).}
\label{fig:tb_3c}
\end{figure}

While the reciprocal space expressions shown in Figs.~\ref{fig:cubic} and \ref{fig:hex} provide insight about spin-splitting patterns due to different types of $\mathcal{I}$-odd orders, they do not give information about the real-space Hamiltonians and hopping integrals directly. 
In this section, we build effective TB models starting from the parent phase $Pm\bar{3}m$ to elucidate the hopping terms that arise from the breaking of $\mathcal{I}$ in each spin-splitting effect. These SOC-induced hopping terms are the real-space counterparts of the SOC Hamiltonian bases $\mathcal{H}_{SOC}(\vb*{k},\vb*{\sigma})$ derived in the previous sections. In order to achieve this, we employ projection operators in real space hopping terms. 
While a complete prescription to obtain the most general TB Hamiltonians for specific symmetries is provided in Ref.~\cite{Kusunose2023}, we take a less general but simpler approach to build models that allow straightforward comparison between different SOC terms by considering atoms on the same crystallographic sites. 
The same approach can also be applied to obtain the $\mathcal{I}$-breaking SOC-induced hopping terms in the $6/mmm$ parent phase, the detailed analysis and results of which are documented in Appendix~\ref{sm:tight binding_hex}.

We consider a single $s$-like orbital on each atom. As such, more than one atom per unit cell is necessary to obtain sufficiently complex models that allow for all irreps. 
In cubic space group $Pm\bar{3}m$, considering an atom in the center of the unit cell (Wyckoff position 1b), along with 3 more atoms on the face centers (Wyckoff position 3c) is sufficient to build models for all NCS-induced SOC splittings except for the Dresselhaus-cubic (D-c) term. This structure is shown in Fig.~\ref{fig:tb_3c}. The discussion of the D-c term requires a unit cell with atoms on Wyckoff positions 1b and 3d (edge-centers) instead. This cell is shown in Fig.~\ref{fig:tb_3d}, and details of the model are presented in Appendix~\ref{sm:D_c}.

We define $c^{\vb*{R}\dagger}_{i\alpha}$ as the creation operator for an electron on atom $0\leq i\leq 3$ in unit cell $\vb*{R}$ with spin $\alpha$. We denote the amplitude of an $n$-dimensional OP $\vb*{F}$ with basis functions $\vb*{F}=[F_1, F_2, ..., F_n]$ as an $n$-component vector $\vb*{f}=[f_1, f_2, ..., f_n]$, and also define the vector of Pauli spin matrices as $\vb*{\sigma}=[\sigma_x, \sigma_y, \sigma_z]$. 
The TB Hamiltonian can be separated into two parts that include a spin-independent part ($\mathcal{H}_0$) and the SOC-induced terms ($\mathcal{H}_{SOC}$) as: 
\begin{equation}
    \mathcal{H}_{tot}=\mathcal{H}_0+\mathcal{H}_{SOC}
\end{equation}
where $\mathcal{H}_0$ can be expressed as: 
\begin{equation}
\label{eqn:H0_1}
    \mathcal{H}_0=\sum_{\vb*{R},\vb*{R'}} \sum_{i,j} t_{ij}^{\vb*{R}\vb*{R'}} \sum_{\alpha, \beta} c_{i,\alpha}^{\vb*{R}\dagger} \hspace{.5mm} \sigma^0_{\alpha \beta} \hspace{.5mm} c_{j,\beta}^{\vb*{R'}} .
\end{equation}
Here, $\sigma^0_{\alpha \beta}=\delta_{\alpha \beta}$ is the identity matrix, and the hopping amplitudes $t_{ij}^{\vb*{RR'}}$ obey all symmetries (including $\mathcal{I}$) of space group $Pm\bar{3}m$. For a minimal model, we consider only the nearest-neighbor hoppings between the atom in the body-center and the atoms on the face-centers. Thus, there is only one hopping parameter $t$ in $\mathcal{H}_0$, which gives: 
\begin{equation}
    \mathcal{H}_0=t\sum_{\vb*{R}} \sum_{j=1}^3 \sum_{\alpha} \left(c_{0,\alpha}^{\vb*{R}\dagger} c_{j,\alpha}^{\vb*{R}}+c_{0,\alpha}^{\vb*{R}\dagger} c_{j,\alpha}^{\vb*{R+a_j}}\right)+h.c.,
\end{equation}
with $\vb*{a_i}~(i\in\{1,2,3\})$ being the lattice vectors of the cubic system along $x,~y,~z$ respectively.

The most general form of $\mathcal{H}_{SOC}$ is: 
\begin{widetext}
\begin{equation}
    \mathcal{H}_{SOC}^\Gamma = 
    i\lambda \sum_{\vb*{R}, \vb*{R'}} \sum_{i,j} \sum_{m} f_m^\Gamma \sum_{n} C_{ij\vb*{R}\vb*{R'}}^{\Gamma,m,n} 
    \left[ \sum_{\alpha, \beta} c_{i,\alpha}^{\vb*{R}\dagger} \hspace{.5mm} \sigma^n_{\alpha \beta} \hspace{.5mm} c^{\vb*{R'}}_{j,\beta}\right] .
    \label{eqn:Hsoc1}
\end{equation}
\end{widetext}
Here, $\lambda$ is proportional to the SOC strength, $f_m^\Gamma$ is the amplitude of the $m$-th component of the order parameter corresponding to an electric multipole that transforms as the $\mathcal{I}$-odd irrep $\Gamma$. $C_{ij\vb*{R R'}}^{\Gamma,m,n}$ are irrep-specific coefficients that determine which hoppings are nonzero. Eq.~\ref{eqn:Hsoc1} can be rewritten in a more concise form by defining: 
\begin{equation}
\label{ac_vec}
    \hat{c}^{\vb*{R}\dagger}_{i}=
    \begin{bmatrix}
    c^{\vb*{R}\dagger}_{i,\uparrow},
    c^{\vb*{R}\dagger}_{i,\downarrow}
    \end{bmatrix},~~ 
    \hat{c}^{\vb*{R}}_{j}=
    \begin{bmatrix}
    c^{\vb*{R}}_{j,\uparrow}\vspace{1mm}\\
    c^{\vb*{R}}_{j,\downarrow}
    \end{bmatrix}.
\end{equation}
This gives  
\begin{equation}
    \sum_{\alpha, \beta} c_{i,\alpha}^{\vb*{R}\dagger} \hspace{.5mm} \sigma^n_{\alpha \beta} \hspace{.5mm} c^{\vb*{R'}}_{j,\beta} \equiv 
    \hat{c}^{\vb*{R}\dagger}_{i} \sigma_n \hat{c}^{\vb*{R'}}_{j} .
\end{equation}
Eq.~\ref{eqn:Hsoc1} then becomes:
\begin{equation}
    \mathcal{H}_{SOC}^\Gamma=i\lambda  \sum_{\vb*{R},\vb*{R'}} \sum_{i,j} \sum_{n} \sum_{m} f_m^\Gamma C_{ij\vb*{R R'}}^{\Gamma,m,n}
    \left(\hat{c}^{\vb*{R}\dagger}_{i} \sigma_n \hat{c}^{\vb*{R'}}_{j}\right).
    \label{eqn:Hsoc1_1}
\end{equation}

Since we consider only the nearest-neighbor hopping terms for $\mathcal{H}_{SOC}$, only $C_{0,j~\vb*{R,R}}^{\Gamma,m,n}$ and $C_{0,j~\vb*{R,R+a_j}}^{\Gamma,m,n}$ (and their hermitian conjugates) are nonzero. An example of such a term is shown in Fig.~\ref{fig:tb_3c}(b). Because the $\Gamma$ irrep is $\mathcal{I}$-odd, the following constraint is imposed:
\begin{equation}
\label{eqn:M}
    \sum_{m} f_m^\Gamma C_{0,j~\vb*{R,R}}^{\Gamma,m,n}
     =-\sum_{m} f_m^\Gamma C_{0,j~\vb*{R,R+a_j}}^{\Gamma,m,n}
    \equiv M_{nj}^\Gamma. 
\end{equation}
Given this constraint, we can further cast Eq.~\ref{eqn:Hsoc1_1} in a simpler form by defining a 3-component vector $\vb*{v^R}$ of $2\times1$ vectors in the spin space: 
\begin{equation}
\label{eqn:tb_vector}
    \vb*{v^R}=\begin{bmatrix}
    v_{1}^{\vb*{R}}
    \vspace{1mm}\\
    v_{2}^{\vb*{R}}
    \vspace{1mm}\\
    v_{3}^{\vb*{R}}
    \end{bmatrix}
    =
    \begin{bmatrix}
        \hat{c}_{1}^{\vb*{R}}-\hat{c}_{1}^{\vb*{R+a_1}}
        \vspace{1mm}\\
        \hat{c}_{2}^{\vb*{R}}-\hat{c}_{2}^{\vb*{R+a_2}}
        \vspace{1mm}\\
        \hat{c}_{3}^{\vb*{R}}-\hat{c}_{3}^{\vb*{R+a_3}}
    \end{bmatrix},
\end{equation}
such that $\mathcal{H}_{SOC}^\Gamma$ can be expressed with the coefficient matrix $M^{\Gamma}$ defined in Eq.~\ref{eqn:M} as:
\begin{equation}
    \mathcal{H}_{SOC}^\Gamma=i\lambda \sum_{\vb*{R}} \sum_{j=1}^3 \sum_{n} M_{nj}^\Gamma\left(\hat{c}_0^{\vb*{R}\dagger}\sigma_n v_j^{\vb*{R}}\right)+h.c..
    \label{eqn:Hsoc2}
\end{equation}

The coefficient matrices $M^\Gamma$ can be determined using the projection operator $\hat{P}^\Gamma$ of each irrep. $\hat{P}^\Gamma$ acting on a spin-orbit Hamiltonian $\mathcal{H}_{SOC}^{\Gamma'}$ gives zero if $\Gamma$ and $\Gamma'$ are not the same irrep, and has the same effect as identity if they are the same: 
\begin{equation} 
    \hat{P}^\Gamma {H}_{SOC}^{\Gamma'}\hat{P}^\Gamma = \delta_{\Gamma \Gamma'} \mathcal{H}_{SOC}^\Gamma .
\end{equation}
Here, $\delta_{\Gamma \Gamma'} = 1$ if and only if $\Gamma$ and $\Gamma'$ are the same irrep, and $\delta_{\Gamma \Gamma'} =0$ otherwise. This constrains the form of $M^\Gamma$ for each irrep, and sets most of its elements to zero. 

\begin{figure*}
\includegraphics[width=0.75\linewidth]{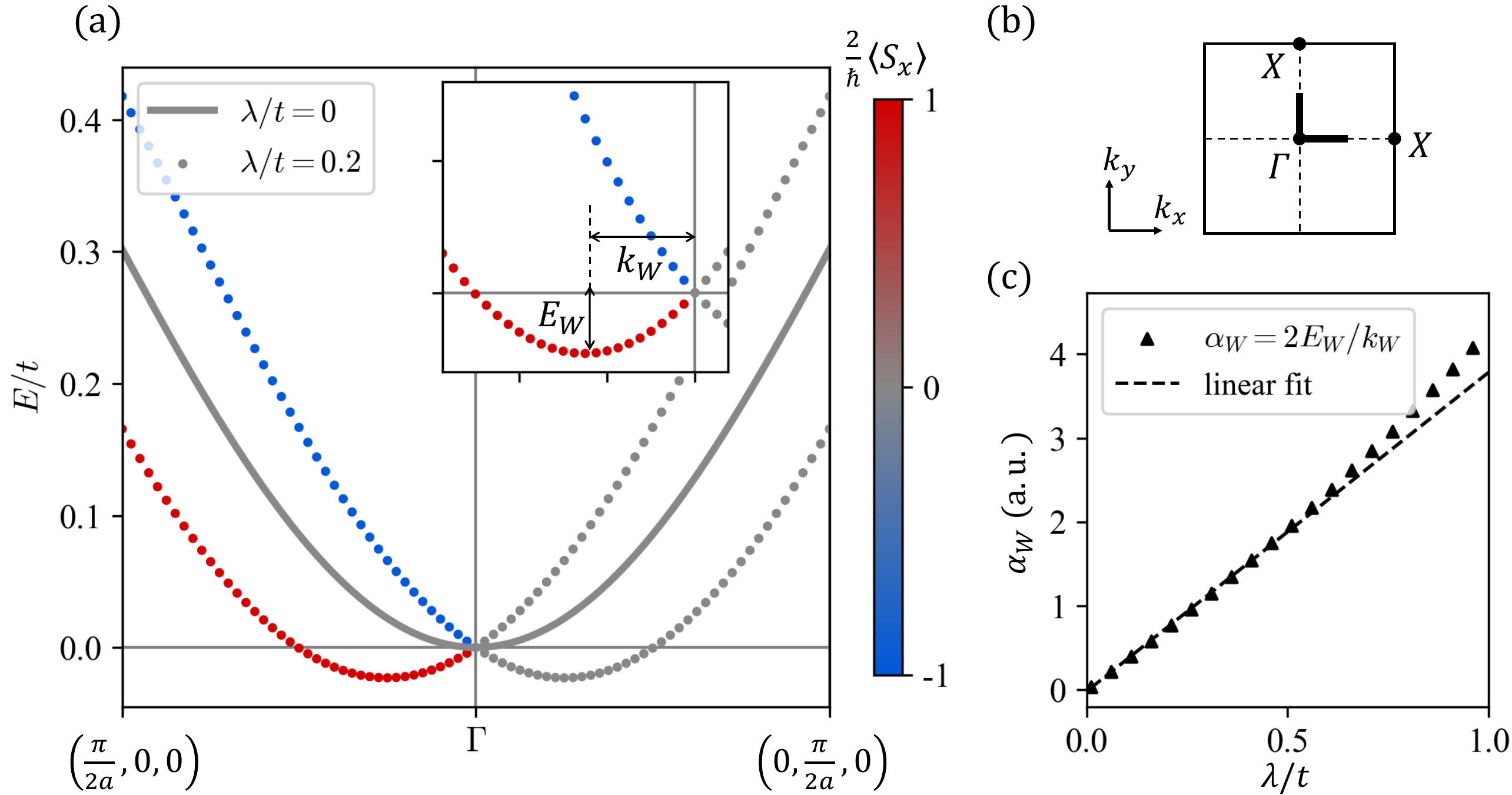}
\caption{(a) Band structure around $\Gamma$ from the TB model with ET-0 (E-512) and SOC hopping strength $\lambda=0$ (solid line) and $\lambda=0.2t$ (dotted line). Color scheme represents the spin expectation value along $x$. The $k$-point path is indicated by thick solid lines in the first Brillouin zone in panel (b). (c) Weyl constant $\alpha_W$ defined as twice the ratio of $E_W$ and $k_W$ (labeled in zoom-in window of (a)) versus relative SOC hopping magnitude $\lambda/t$. Highly linear behavior is observed around $\lambda\rightarrow0$.}
\label{fig:ETM-alpha}
\end{figure*}

How $\hat{P}^\Gamma$ acts on the TB Hamiltonians can be understood from how crystallographic operations act on the atomic positions and on $\vb*{\sigma}$, which transforms as an axial vector in the spin space (basis~=~$\hat{c}^{\vb*{R}\dagger}_{i}$). For example, a 90$^\circ$ rotation around the $z$-axis going through the center atom in Fig.~\ref{fig:tb_3c}, which we denote as $4^+_{001}$, moves an electron on site 1 in the unit cell $\vb*{R}$ to the site 2 in the same unit cell, while leaving the electron on site 0 where it was. 
In other words, $4_{001}^+ c_{0\alpha}^{\vb*{R}\dagger} 4_{001}^- = e^{-i\frac{\pi}{4}\alpha}c_{0\alpha}^{\vb*{R}\dagger}$ 
but $4_{001}^+ c_{1\alpha}^{\vb*{R}\dagger} 4_{001}^- = e^{-i\frac{\pi}{4}\alpha}c_{2\alpha}^{\vb*{R}\dagger}$ with $\alpha=\pm 1$ for up and down spins. Here, the spin-dependent phase $e^{-i\frac{\pi}{4}\alpha}$ comes in from the spin-$1/2$ nature of the operators. This gives, for example:
\begin{equation}
\begin{aligned}
4^+_{001} \left(\hat{c}_{0}^{\vb*{R}\dagger}\sigma_x~\hat{c}_{1}^{\vb*{R}}\right) 4^-_{001}
    & = 4^+_{001} \left(c_{0\uparrow}^{\vb*{R}\dagger}c_{1\downarrow}^{\vb*{R}}+c_{0\downarrow}^{\vb*{R}\dagger}c_{1\uparrow}^{\vb*{R}}\right) 4^-_{001}\\
    & = -ic_{0\uparrow}^{\vb*{R}\dagger}c_{2\downarrow}^{\vb*{R}}+ic_{0\downarrow}^{\vb*{R}\dagger}c_{2\uparrow}^{\vb*{R}}\\
   & =\hat{c}_{0}^{\vb*{R}\dagger} \sigma_y \hat{c}_{2}^{\vb*{R}}.
\end{aligned}
\end{equation}
The projection operators can be built by considering the effect of all crystallographic operations with coefficients determined by the characters of different irreps. Using these operators, we obtain the TB models corresponding to each $\mathcal{I}$-odd irrep of the cubic group, which are summarized in Fig.~\ref{fig:cubic}(c) in the form of $M$ matrices. The $M$ matrices have forms reminiscent of rank-2 tensor components with the same irreps that the multipoles correspond to. While we do not present a formal proof, one can explain this observation by associating the matrix $M$ with the Jahn symbol $eV^2$. The Rashba (R) term gives rise to an antisymmetric $M$ with three components that transform as $T_{1u}^+$, whereas the Dresselhaus (D-1 and D-2) terms lead to traceless symmetric $M$'s with components that transform as $T_{2u}^+$ and $E_u^+$ irreps. Finally, the Weyl (W) term has an $M$ matrix with a single component that transforms as $A_{1u}^+$. The sum of these irreps gives the representation of the $eV^2$ tensor according to the point group tables~\cite{Aroyo2006}: $eV^2\rightarrow A_{1u}^+ + E_u^+ + T_{1u}^+ + T_{2u}^+$. 

Band structures of TB models constructed from the unit cell in Fig.~\ref{fig:tb_3c} reproduce the same spin patterns near the zone center as the linear-in-$\vb*{k}$ SOC Hamiltonians we discussed earlier (Fig.~\ref{fig:cubic}(a,c)), not including the Dresselhaus cubic (D-c) term, whose lowest-order $\mathcal{H}_{SOC}$ is cubic in $\vb*{k}$. All types of SOC terms in the TB models discussed above give rise to band splittings starting from the linear order in $\vb*{k}$. As a result, the spin-degenerate parabolic band in the parent $m\bar{3}m$ phase splits into two subbands with opposite spin states and linear energetic dispersion with opposite slope in the vicinity of $\Gamma$, described by $E=C(k^2\pm\alpha_{soc} k)$, where $C$ is some constant and $\alpha_{soc}$ is the SOC coefficient that quantifies, in each case, the magnitude of the spin-splitting. 

Fig.~\ref{fig:ETM-alpha} shows the Weyl-type spin-splitting induced by the emergence of ET-0 (E-512) in the TB model. The Weyl coefficient $\alpha_W$ can be calculated as $\alpha_W= 2E_W/k_W$, where the Weyl energy $E_W$ and Weyl momentum $k_W$ are defined in the inset of panel (a). Panel (c) reveals a linear relationship between $\alpha_W$ and the SOC hopping amplitude $\lambda$ for small $\lambda$ with respect to $t$, as expected. We confirmed the linear relationship for other SOC types as well (not shown). This can be considered as a generalization of the linear relationship between the Rashba coefficient $\alpha_R$ and the magnitude of the electric dipole (E-1) in polar systems~\cite{Leppert2016,Guo2023rashba,Yao2017}. By controlling different structural orders via strain~\cite{Tao2016,Tyunina2010}, layering~\cite{Shanavas2014,Yin2018}, or other methods, this observation can provide alternative means to control SOC Hamiltonians in solids to, for example, stabilize persistent spin helices~\cite{Bernevig2006, Koralek2009, Tao2018}.

\section{Secondary Orders and Spin-Splitting Nodal Structures}
\label{sec:secondary}

\begin{table}
\renewcommand{\arraystretch}{1.5}
\caption{Secondary OPs induced by primary $\mathcal{I}$-odd OPs of the $m\bar{3}m$ parent group and the corresponding nodal structures in the vicinity of the $\Gamma$-point of the Brillouin zone.}

\begin{tabular}{ 
  | >{\centering\arraybackslash}p{0.08\linewidth}  
  | >{\centering\arraybackslash}p{0.08\linewidth}   
  | >{\centering\arraybackslash}p{0.14\linewidth}   
  | >{\centering\arraybackslash}p{0.2\linewidth}   
  | >{\centering\arraybackslash}p{0.18\linewidth}   
  | >{\centering\arraybackslash}p{0.2\linewidth}   | }
    \hline
     \multicolumn{3}{| >{\centering\arraybackslash}p{0.30\linewidth} |}{Primary OP} & Secondary OP & Subgroup & Nodal structure\\
    \hline
    \multirow{6}{*}{R} &\multirow{6}{*}{$T_{1u}^+$} & (a,0,0) & none & $4mm$ & [100] line \\ 
    \cline{3-6}  
    & & (a,a,0) & $T_{2u}^+$(a,-a,0) & $mm2$ & [110] line \\
    \cline{3-6}  
    & & (a,a,a) & $A_{2u}^+$(a) & $3m$ & [111] line \\
    \cline{3-6}  
    & & (a,b,0) & $T_{2u}^+$(a,b,0) & $m^{\dagger}$ & point, line$^*$ \\
    \cline{3-6}  
    & & (a,a,b) & $A_{2u}^+$(a), $E_{u}^+$(a,0), $T_{2u}^+$(a,-a,0) & $m^{\ddagger}$ & point, line$^*$ \\
    \cline{3-6}  
    & & (a,b,c) & all & $1$ & point \\
    \hline
    \multirow{6}{*}{D-1} &\multirow{6}{*}{$T_{2u}^+$} & (a,0,0) & none & $\bar{4}m2$ & [100] line \\ 
    \cline{3-6}  
    & & (a,a,0) & $T_{1u}^+$(a,-a,0) & $mm2$ & $[1\bar{1}0]$ line \\
    \cline{3-6}  
    & & (a,a,a) & $A_{1u}^+$(a) & $32$ & point\\
    \cline{3-6}  
    & & (a,b,0) & $T_{1u}^+$(a,b,0) & $m^{\dagger}$ & point, line$^*$\\
    \cline{3-6}  
    & & (a,a,b) & $A_{1u}^+$(a), $E_{u}^+$(0,a), $T_{1u}^+$(a,-a,0) & $2$ & point \\
    \cline{3-6}  
    & & (a,b,c) & all & $1$ & point \\
    \hline
    \multirow{3}{*}{D-2} &\multirow{3}{*}{$E_{u}^+$} & (a,0) & $A_{2u}^+$(a) & $\bar{4}2m$ & [001] line \\ 
    \cline{3-6}  
    & & (0,a) & $A_{1u}^+$(a) & $422$ & point \\
    \cline{3-6}  
    & & (a,b) & $A_{1u}^+$(a), $A_{2u}^+$(a) & $222$ & point \\
    \hline 
    W &$A_{1u}^+$ & (a) & none & $432$ & point \\
    \hline
    D-c &$A_{2u}^+$ & (a) & none & $\bar{4}3m$ & $\langle100\rangle$ \& $\langle111\rangle$ lines \\
    \hline   
\end{tabular}
\label{table:nodes}
\raggedright
\footnotesize{$^*$ Accidental degeneracy}\\
\footnotesize{$^{\dagger}$ $m\perp[001]$ ~$^{\ddagger}$ $m\perp[110]$} 
\end{table}

\begin{table}
\renewcommand{\arraystretch}{1.5}
\caption{Secondary OPs induced by primary $\mathcal{I}$-odd OPs of the $6/mmm$ parent group and the corresponding nodal structures in the vicinity of the $\Gamma$-point of the Brillouin zone.}
\begin{tabular}{ 
  | >{\centering\arraybackslash}p{0.08\linewidth}  
  | >{\centering\arraybackslash}p{0.08\linewidth}   
  | >{\centering\arraybackslash}p{0.14\linewidth}   
  | >{\centering\arraybackslash}p{0.2\linewidth}   
  | >{\centering\arraybackslash}p{0.18\linewidth}   
  | >{\centering\arraybackslash}p{0.2\linewidth}   | }
\hline
\multicolumn{3}{| >{\centering\arraybackslash}p{0.30\linewidth}|}{Primary OP} & Secondary OP & Subgroup & Nodal structure\\
\hline
R & $A_{2u}^+$ & (a) & none & $6mm$ & [001] line  \\
\hline
\multirow{3}{*}{D} & \multirow{3}{*}{$E_{2u}^+$} & (a,0) & $A_{2u}^+$(a) & $mm2^\dagger$ & [001] line\\ 
\cline{3-6}  
& & (0,a) & $A_{1u}^+$(a) & $222$ & point \\
\cline{3-6}  
& & (a,b) & $A_{1u}^+$(a), $A_{2u}^+$(a)  & $2$ & point \\
\cline{3-6}  
\hline
\multirow{3}{*}{R-D} & \multirow{3}{*}{$E_{1u}^+$} & (a,0) & $B_{2u}^+$(a) & $mm2^\dagger$ & [100] line \\ 
\cline{3-6}  
& & (0,a) & $B_{1u}^+$(a) & $mm2^\dagger$ & [120] line \\
\cline{3-6}  
& & (a,b) & $B_{1u}^+$(a), $B_{2u}^+$(a)  & $m$ & point, line$^*$ \\
\hline
W & $A_{1u}^+$ & (a) & none & $622$ & point  \\
\hline
I-1 & $B_{1u}^+$ & (a) & none & $\bar{6}$m2 & [001] \& $\langle120\rangle$ lines \\
\hline
I-2 & $B_{2u}^+$ & (a) & none & $\bar{6}2m$ & [001] \& $\langle100\rangle$ lines \\
\hline
\end{tabular}
\label{table:nodes_hex}
\\
\raggedright
\footnotesize{$^*$ Accidental degeneracy} \\ 
\footnotesize{$^\dagger$ 2-axes (i.e., intersecting lines of mirrors) are along different directions in each case.}   
\end{table}

Degeneracies between otherwise spin-split bands in NCS materials, including the nodal points (e.g., Weyl nodes) and lines, have been studied in detail, especially from the point of view of topology (e.g., Ref.~\cite{Burkov2011, Yang2018, Armitage2018, Xie2021, Samokhin2009}). However, these studies mostly focus on different point groups without distinguishing different types of SOC terms and ways to break $\mathcal{I}$ using the irreps of a reference group. In this section, we approach this problem using irreps and elucidate the effect of secondary orders, which are OPs that are necessarily nonzero in the presence of a primary order with a different irrep. 

In Fig.~\ref{fig:pg_venn}, we showed a Venn diagram with the allowed SOC effects in all NCS point groups. Each of these groups can be obtained by breaking the symmetries of either $m\bar{3}m$ or $6/mmm$ via the condensation of one or more $\mathcal{I}$-odd irreps, as shown in Fig.~\ref{fig:pg_tree}. These diagrams make it explicit that most NCS groups host a combination of R, D, W, or I type of SOC terms, rather than a single one of them. 

The reason why multiple SOC terms that transform as different irreps often coexist in NCS systems is not only because the symmetry breaking path towards the corresponding subgroup requires multiple irreps. Instead, it is often the case that a single primary OP driving a phase transition breaks enough symmetries to induce secondary OPs with different irreps~\cite{Toledano1987}. This is only possible for OPs that transform as multi-dimensional irreps, since powers of a one-dimensional irrep transform either as the irrep itself (odd powers) or as the trivial irrep (even powers). A well-known example of the emergence of secondary OPs is that most structural phase transitions driven by an unstable phonon mode, such as ferroelectricity in BaTiO$_3$, also induce a strain that is proportional to the square of the primary OP's amplitude. Interestingly, these secondary OPs can, in certain conditions, condense before the primary OP condenses, giving rise to a so-called vestigial phase \cite{Fernandes2019}. In Table~\ref{table:nodes}, we list all $\mathcal{I}$-odd secondary OPs induced by different directions of primary OPs. These secondary OPs can be found by looking for higher-order terms in the Landau free-energy expansion of the system, as secondary OPs are linearly coupled to powers of the corresponding primary OPs. One can also consider the point groups obtained by the condensation of the primary OP: Any other irrep that leads to a supergroup of this point group would necessarily be present as a secondary irrep in the subgroup. 

\begin{figure}
\includegraphics[width=\linewidth]{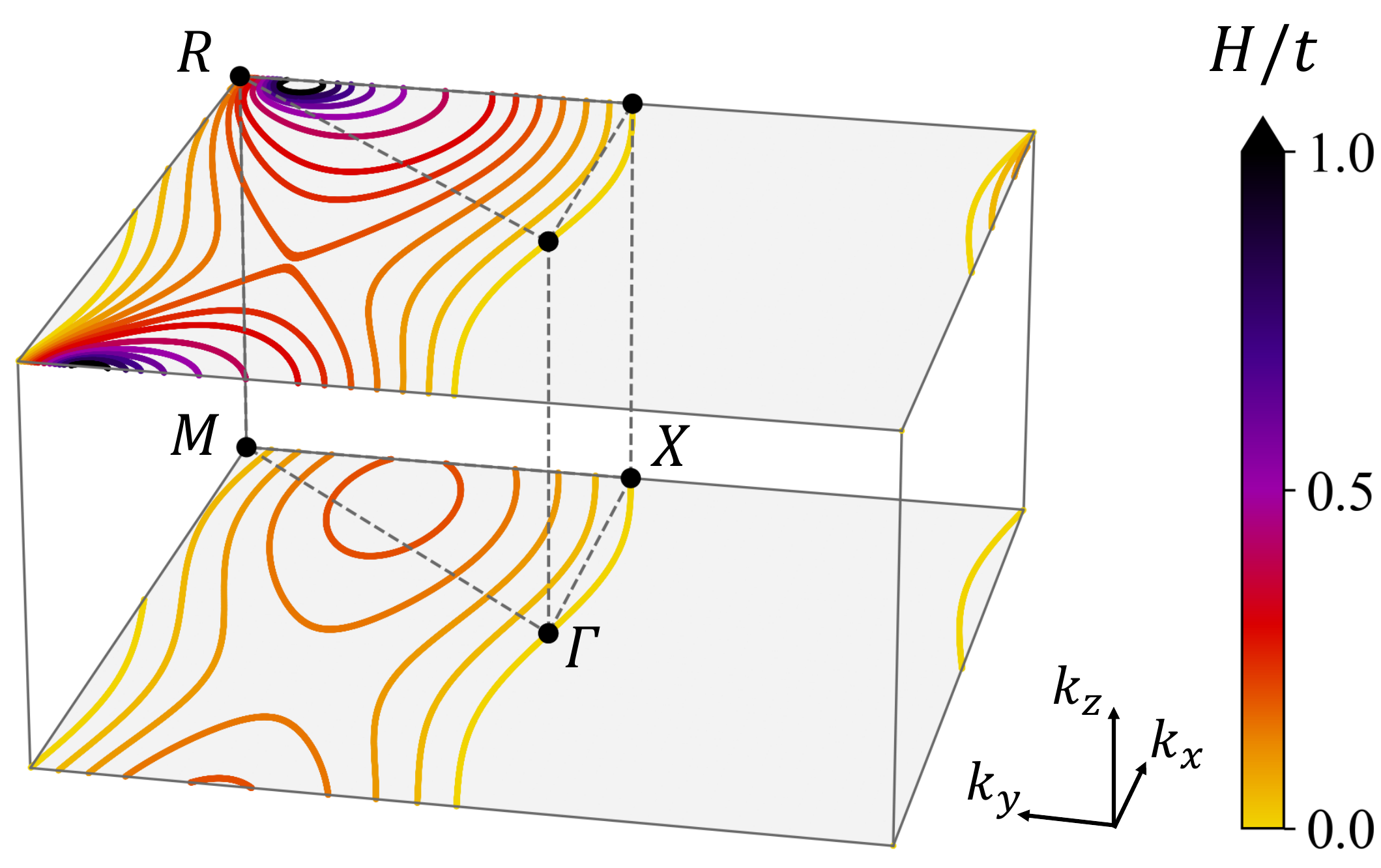}
\caption{Evolution of the nodal structure as a function of Zeeman splitting $H$ (indicated by color scheme) in the two lowest bands of the TB model with SOC terms corresponding to the primary OP $T_{1u}^+(0.2,0.4,0)$ and the secondary OP $T_{2u}^+(0.3,0.5,0)$ in the cubic parent phase $m\bar{3}m$. For small values of $H$, there are two nodal lines on each mirror-invariant plane, $k_z=0$ and $k_z=\pi/a$. as $H$ increases, the pairs get closer, merge to form nodal loops, and finally disappear. }
\label{fig:nodal}
\end{figure}

Most $\mathcal{I}$-breaking orders in $m\bar{3}m$ lead to one or more secondary orders, since most $\mathcal{I}$-odd OPs are multi-dimensional. Applying the TB or reciprocal space models derived above without considering these secondary effects may lead to band degeneracies (i.e., spin-splitting nodes) that are not present in the material. For example, for a polar order $\vb*{P}$, the Hamiltonian near the zone-center can be written as $\mathcal{H}=\frac{k^2}{2m}\sigma_0 + \alpha_R \vb*{\sigma}\cdot (\vb*{P}\times \vb*{k})$, with $\sigma_0$ being the $2\times2$ unit matrix. This Hamiltonian always gives rise to a nodal line (i.e. absence of spin-splitting) parallel to $\vb*{P}$. However, if $\vb*{P}$ is along a low-symmetry direction $(a,b,c)$ with $a\neq b\neq c$, then according to Table~\ref{table:nodes}, all other $\mathcal{I}$-breaking irreps are also condensed as secondary OPs. Among these is $A_{1u}^+$ (the Weyl term), which only allows for a spin-splitting node at the $\Gamma$ point. Hence, when the Rashba term is induced in a material through a polar order along a low-symmetry direction $(a,b,c)$, the other SOC effects need to be taken into account as secondary orders to obtain the correct nodal structure, which in this case consists of a single nodal point. 

In the last column of Table~\ref{table:nodes}, we list symmetry-imposed nodal points and lines around the zone center that emerge from different directions of $\mathcal{I}$-odd primary OPs, taking into account the secondary OPs as well. A similar table (Table~\ref{table:nodes_hex}) is shown for the case of $\mathcal{I}$-odd orders originating from the parent hexagonal group $6/mmm$. 
Together, our results are consistent with earlier studies which provided similar lists based on NCS point groups only~\cite{Samokhin2009, Xie2021}\footnote{The only disagreement between our work and Ref.~\cite{Samokhin2009} is that we predict a nodal line of accidental degeneracies for the point group $m$, in line with Ref.~\cite{Xie2021}.}. In particular, we confirm the observation in Ref.~\cite{Xie2021} that as long as the point group is not chiral, every NCS point group hosts nodal lines (For point groups not mentioned in Table~\ref{table:nodes} and \ref{table:nodes_hex}, see Appendix~\ref{sm:nodal}). In our approach, this can be understood from the fact that as long as $A_{1u}^+$, which corresponds to an electrotoroidal monopole (ET-0) and hence chirality~\cite{Inda2024, Kusunose2024}, is absent, nodal lines are always present. 

\begin{table*}
\renewcommand{\arraystretch}{1.5}
\centering
\caption{NCS space groups (trigonal and hexagonal systems excluded), experimentally synthesized materials examples, and allowed spin-splitting terms listed according to their point groups.}
\begin{tabular}{ 
  | >{\centering\arraybackslash}p{0.08\linewidth}  
  | >{\centering\arraybackslash}p{0.33\linewidth}     
  | >{\centering\arraybackslash}p{0.05\linewidth}   
  | >{\centering\arraybackslash}p{0.05\linewidth}   
  | >{\centering\arraybackslash}p{0.05\linewidth}   
  | >{\centering\arraybackslash}p{0.05\linewidth} 
  | >{\centering\arraybackslash}p{0.05\linewidth} 
  | >{\centering\arraybackslash}p{0.26\linewidth}   | }
    \hline
        NCS PG & NCS SG & R & D-1 & D-2 & W & D-c & Materials Examples \\
    \hline
        $1$ & $P1$ & \cmark & \cmark & \cmark & \cmark & \cmark & N/A\\
    \hline
        $2$ & $P2$, $P2_1$, $C2$ & \cmark & \cmark & \cmark & \cmark & \cmark & CsP(HO$_2$)$_2$, Eu$_2$B$_5$Os$_3$, Ca$_2$B$_5$Os$_3$\\
    \hline
        $m$ & $Pm$, $Pc$, $Cm$, $Cc$ & \cmark & \cmark & \cmark & \xmark & \cmark & Ta$_3$SBr$_7$, AgSbS$_2$, LiNdTi$_2$O$_6$\\
    \hline
        $222$ & $P222$, $P222_1$, $P2_12_12$, $P2_12_12_1$, $C222_1$, $C222$, $F222$, $I222$, $I2_12_12_1$ & \xmark & \cmark & \cmark & \cmark & \cmark & AlPS$_4$, MgTeMoO$_6$, CdOF\\
    \hline
        $mm2$ & $Pmm2$, $Pmc2_1$, $Pcc2$, $Pma2$, $Pca2_1$, $Pnc2$, $Pmn2_1$, $Pba2$, $Pna2_1$, $Pnn2$, $Cmm2$, $Cmc2_1$, $Ccc2$, $Amm2$, $Aem2$, $Ama2$, $Aea2$, $Fmm2$, $Fdd2$, $Imm2$, $Iba2$, $Ima2$ & \cmark & \cmark & \cmark & \xmark & \cmark & Ba$_2$In$_2$O$_5$, SrIn, KNb(AgSe$_2$)$_2$\\
    \hline
        $4$ & $P4$, $P4_1$, $P4_2$, $P4_3$, $I4$, $I4_1$, $P\bar{4}$, $I\bar{4}$, $P4/m$, $P4_2/m$, $P4/n$, $P4_2/n$, $I4/m$, $I4_1/a$ & \cmark & \xmark & \cmark & \cmark & \xmark & Cr$_4$AgBiO$_{14}$, WBr$_4$O, IrN$_7$ClO$_6$\\
    \hline
        $422$ & $P422$, $P42_12$, $P4_122$, $P4_12_12$, $P4_222$, $P4_22_12$, $P4_322$, $P4_32_12$, $I422$, $I4_122$ & \xmark & \xmark & \cmark & \cmark & \xmark & TeO$_2$, CdAs$_2$, TiPbO$_3$\\
    \hline
        $4mm$ & $P4mm$, $P4bm$, $P4_2cm$, $P4_2nm$, $P4cc$, $P4nc$, $P4_2mc$, $P4_2bc$, $I4mm$, $I4cm$, $I4_1md$, $I4_1cd$ & \cmark & \xmark & \xmark & \xmark & \xmark & KNbO$_3$, CeAl$_3$Au, TaAs\\
    \hline
        $\bar{4}$ & $P\bar{4}$, $I\bar{4}$ & \xmark & \cmark & \cmark & \xmark & \cmark & AgC$_8$S$_4$N$_4$Cl, Cd(InTe$_2$)$_2$, TbCd$_2$F$_8$\\
    \hline
        $\bar{4}m2$ & $P\bar{4}m2$, $P\bar{4}c2$, $P\bar{4}b2$, $P\bar{4}n2$, $I\bar{4}m2$, $I\bar{4}c2$ & \xmark & \cmark & \xmark & \xmark & \xmark & Hf$_2$Sb$_3$Pd, Sn$_3$Ru$_2$, RbMnTe$_2$\\
    \hline
        $\bar{4}2m$ & $P\bar{4}2m$, $P\bar{4}2c$, $P\bar{4}2_1m$, $P\bar{4}2_1c$, $I\bar{4}2m$, $I\bar{4}2d$ & \xmark & \xmark & \cmark & \xmark & \cmark & Ag$_2$HgI$_4$, AlPS$_4$, Pd$_4$Se\\
    \hline
        $23$ & $P23$, $F23$, $I23$, $P2_13$, $I2_13$ & \xmark & \xmark & \xmark & \cmark & \cmark & AlAu$_4$, Bi$_2$Pd$_3$S$_2$, SrSiPt\\
    \hline
        $432$ & $P432$, $P4_232$, $F432$, $F4_132$, $I432$, $P4_332$, $P4_132$, $I4_132$ & \xmark & \xmark & \xmark & \cmark & \xmark & AsH$_3$, BaSi$_2$, Nb$_3$Au$_2$, \\
    \hline
        $\bar{4}3m$ & $P\bar{4}3m$, $F\bar{4}3m$, $I\bar{4}3m$, $P\bar{4}3n$, $F\bar{4}3c$, $I\bar{4}3d$ & \xmark & \xmark & \xmark & \xmark & \cmark & ZnO, Ag$_3$AsO$_4$, Ba$_4$Bi$_3$\\
    \hline 
\end{tabular}
\label{table:materials_cubic}
\end{table*}

In addition to distinguishing whether a NCS material has symmetry-imposed nodal points (Kramers-Weyl nodes) or nodal lines (Kramers-Weyl nodal lines)~\cite{Xie2021,He2021,Zhang2023}, one can further distinguish between different types of symmetry-imposed nodal lines. In any NCS point group, opposite-spin bands are always degenerate at the $\Gamma$ point (as well as the other TRIM points) because of $\mathcal{T}$. This degeneracy may or may not be separately protected by crystalline symmetries as well. For example, in the point group $4mm$, all fermionic irreps of the double group are at least 2-dimensional, which means that opposite spin electrons are forced to be degenerate by rotation or mirror symmetries at the $\Gamma$ point even in the absence of $\mathcal{T}$. The same 2-dimensional irreps also exist along the entire [001] line, ensuring that there is a crystalline symmetry-protected nodal line along this axis. On the other hand, in some lower-symmetry double point groups, e.g., $\bar{4}$ ($\bar{6}$) and $m$, there are only 1-dimensional fermionic irreps, where the degeneracy at $\Gamma$ is purely a result of Kramers' theorem. However, in $\bar{4}$ ($\bar{6}$), these irreps ``stick together'' along the whole improper rotational axis under $\mathcal{T}$; while in $m$, these irreps only ``stick together'' at the $\Gamma$ point~\cite{Dresselhaus2007, Lax2001}. In the latter case, a nodal line that extends from $\Gamma$ along a low-symmetry path on the mirror plane consists of accidental degeneracies, as there is no symmetry to impose spin degeneracy at an arbitrary $\vb*{k}$-point. For example, in the case where there are nonzero $T_{1u}^+(p_x,p_y,0)$ and $T_{2u}^+(d_x,d_y,0)$ orders, the spin-splitting $\mathcal{H}_{SOC}(\vb*{k},\vb*{\sigma})$ of a single pair of bands near $\Gamma$ is
\begin{equation}
\mathcal{H}_{SOC}=\left[\alpha_R(p_xk_y - p_yk_x) + \alpha_D (d_x k_y+d_yk_x)\right] \sigma_z, 
\end{equation} 
which vanishes when 
\begin{equation}
    k_y = \frac{\alpha_R p_y -\alpha_D d_y}{\alpha_R p_x+\alpha_D d_x} k_x .
\end{equation}
In Tables~\ref{table:nodes} and \ref{table:nodes_hex}, we list nodal lines like this, which consist of accidental degeneracies between bands with different irreps but is always present in certain point groups because of $\mathcal{T}$, with an asterisk superscript. We underline that these are different from accidental nodal features that may or may not exist depending on parameters of the electronic Hamiltonian. 

The same result can also be obtained from the TB models as well. In Fig.~\ref{fig:nodal}, we show the nodal lines predicted by a TB model with both $T_{1u}^+(p_x,p_y,0)$ and $T_{2u}^+(d_x,d_y,0)$ terms in yellow. There are four nodal lines, each connecting a pair of $\mathcal{T}$-invariant momenta. 

Interestingly, when a magnetic field $H$ normal to the mirror plane ($z$-direction) is applied, the nodal lines do not disappear for small $H$. We model this by adding an onsite Zeeman term $\propto H \sigma_z$ to the TB model, and display the results for varying strengths of $H$ in Fig.~\ref{fig:nodal}. Such a magnetic field only breaks $\mathcal{T}$ and preserves the $m$ plane. As a result, $\mathcal{T}$ is no longer present to pin the nodal lines to TRIM points, and for small values of $H$, the nodal lines are displaced but still intact. With increasing $H$, pairs of nodal lines merge and form nodal loops. Larger values of $H$ shrink these loops and destroy them. As a result, under the magnetic field, there are series of topological transitions between different nodal configurations, very much similar to the topological transitions predicted in altermagnets under the effect of magnetic fields~\cite{Fernandes2024}. The key difference between the two types of systems is that in the present case, the $H$ field only breaks $\mathcal{T}$, whereas in altermagnets, it breaks crystalline symmetries (or their combinations with $\mathcal{T}$) as well. In either case, a small field does not destroy the entire nodal structure, and the transitions where nodes disappear are between phases that are isostructural. A similar situation happens on the single out-of-plane mirror in point group $\bar{6}$ (Fig.~\ref{fig:nodal_hex}), where, between the two lowest spin-polarized bands, there are three symmetry-imposed ``accidental'' nodal lines crossing at $\Gamma$($A$) on each mirror-invariant plane when external magnetic field is absent.

Finally, in order to illustrate that these models are realized in many materials and provide examples of materials of interest, we list some representative examples of materials in all NCS point groups in Table~\ref{table:materials_cubic} for subgroups of $m\bar{3}m$, and Table~\ref{table:materials_hex} for subgroups of $6/mmm$. To obtain these examples, we performed a materials database search and generated a list of experimentally observed materials in the Materials Project Database~\cite{Jain2013}. Even though some space groups are not reported in any materials in this database, every NCS point group is realized in multiple examples. 

\begin{table*}
\renewcommand{\arraystretch}{1.5}
\centering
\caption{NCS space groups in the trigonal and hexagonal systems, experimentally synthesized materials examples, and allowed spin-splitting terms listed according to their point groups.}
\begin{tabular}{ 
  | >{\centering\arraybackslash}p{0.08\linewidth}  
  | >{\centering\arraybackslash}p{0.22\linewidth}   
  | >{\centering\arraybackslash}p{0.05\linewidth}   
  | >{\centering\arraybackslash}p{0.05\linewidth}   
  | >{\centering\arraybackslash}p{0.05\linewidth}   
  | >{\centering\arraybackslash}p{0.05\linewidth}   
  | >{\centering\arraybackslash}p{0.05\linewidth} 
  | >{\centering\arraybackslash}p{0.05\linewidth} 
  | >{\centering\arraybackslash}p{0.30\linewidth}   | }
    \hline
        NCS PG & NCS SG & R & D & R-D & W & I-1 & I-2 & Materials Examples \\
    \hline
        $3$ & $P3$, $P3_1$, $P3_2$, $R3$ & \cmark & \xmark & \xmark & \cmark & \cmark & \cmark & FeBiO$_3$, InNi$_2$SbO$_6$\\
    \hline
        $312$ & $P312$, $P3_112$, $P3_212$ & \xmark & \xmark & \xmark & \cmark & \cmark & \xmark & NaNiIO$_6$, KMnAg$_3$(CN)$_6$\\
    \hline
        $321$ & $P321$, $P3_121$, $P3_221$, $R32$ & \xmark & \xmark & \xmark & \cmark & \xmark & \cmark & Cs$_2$TeO$_3$, BaAlB$_2$O$_7$, NdAl$_3$(BO$_3$)$_4$\\
    \hline
        $31m$ & $P31m$, $P31c$ & \cmark & \xmark & \xmark & \xmark & \xmark & \cmark & Cs$_3$As$_5$O$_9$, KAlSiO$_4$, RbLiCrO$_4$\\
    \hline
        $3m1$ & $P3m1$,$P3c1$, $R3m$, $R3c$ & \cmark & \xmark & \xmark & \xmark & \cmark & \xmark & Mn$_2$NiO$_{14}$, Cu$_3$TeS$_3$Cl, KNbO$_3$\\
    \hline
        $6$ & $P6$, $P6_1$, $P6_2$, $P6_3$, $P6_4$, $P6_5$ & \cmark & \xmark & \xmark & \cmark & \xmark & \xmark & Ag$_3$AsSe$_3$, KAlSiO$_4$, RbLiCrO$_4$\\
    \hline
        $622$ & $P622$, $P6_122$, $P6_222$, $P6_322$, $P6_422$, $P6_522$  & \xmark & \xmark & \xmark & \cmark & \xmark & \xmark &  CePO$_4$, AlPO$_4$, Nb$_3$NiSe$_6$\\
    \hline
        $6mm$ & $P6mm$, $P6cc$, $P6_3cm$, $P6_3mc$  & \cmark & \xmark & \xmark & \xmark & \xmark & \xmark & CsV$_2$O$_5$, NdAgPb, CsCuCl$_3$\\
    \hline
        $\bar{6}$ & $P\bar{6}$ & \xmark & \xmark & \xmark & \xmark & \cmark & \cmark &  Er$_3$(In$_2$Co)$_2$, Yb(P$_2$Rh$_3$)$_2$ \\
    \hline
        $\bar{6}m2$ & $P\bar{6}m2$, $P\bar{6}c2$ & \xmark & \xmark & \xmark & \xmark & \cmark & \xmark & BaSr(FeO$_2$)$_4$, Ca(As$_2$Rh$_3$)$_2$, LiScI$_3$\\
    \hline
        $\bar{6}2m$ & $P\bar{6}2m$, $P\bar{6}2c$ & \xmark & \xmark & \xmark & \xmark & \xmark & \cmark &  LiTbGe, GdInPt, Yb$_3$Ge$_5$, ZrSnIr\\
    \hline  
    \end{tabular}
\label{table:materials_hex}
\end{table*}

\section{Summary and Conclusions}
\label{sec:conc}

In this study, we considered the SOC-induced spin-splitting in the electronic bands of $\mathcal{T}$-invariant crystals using a symmetry-based approach. Utilizing irrep projection operators, we obtained not only $\vb*{k}$-space Hamiltonians but also TB models that allow studying different types of $\mathcal{I}$-breaking orders within the same framework. Using irreps of only two high-symmetry point groups (cubic $m\bar{3}m$ and hexagonbal $6/mmm$), we classified all distinct SOC terms allowed in NCS crystallographic point groups, without the need to go through all the 21 NCS point groups individually. 
This procedure also showed that any SOC terms in a crystallographic solid can be represented as a mixture of four different types of SOC effects: Rashba, Dresselhaus, Weyl, and Ising types. 

Moreover, we derived all $\mathcal{I}$-breaking secondary orders that are induced by primary orders, and tabulated their combined effects on the nodal structures of the spin-split bands. This enabled distinguishing different types of nodal lines in NCS crystals~\cite{He2021}, and showing how to manipulate them (e.g., by a magnetic field), leading to topological spin-splitting transitions similar to those in altermagnets~\cite{Fernandes2024}. 

In addition to providing a unified and simplified approach to SOC in crystallographic solids, and providing real materials examples of where such effects are realized, our study connects to the general understanding of spin-splitting phenomena. Generally, it is the combination of $\mathcal{T}$ and $\mathcal{I}$ symmetries that ensure spin-degenerate bands in a material. There are different ways by which these symmetries can be broken, where one case is to break $\mathcal{T}$ and preserve $\mathcal{I}$ while turning \textit{off} the effects of the SOC. Beyond the trivial case of ferromagnets, where the spin-splitting is essentially uniform in momentum space, this case also encompasses the wide class of altermagnets  \cite{Smejkal2022_1,Smejkal2022_2,Jungwirth2026}. These magnetically-ordered systems break $\mathcal{T}$ but are invariant under a combination of $\mathcal{T}$ with a point group crystalline operation that is not $\mathcal{I}$ (such as rotation), resulting in even-parity spin-splitting. 

Our analysis in this paper focuses on the opposite limiting case where  $\mathcal{T}$ is preserved while $\mathcal{I}$ is broken, and the effects of SOC are essential to give an odd-parity spin-splitting. Nevertheless, there is an interesting analogy with altermagnetism, particularly for the subset of NCS systems that are described by a higher-order electric multipole moment (i.e., beyond dipolar), corresponding to the Dresselhaus, Weyl, and Ising types of SOC effects in Figs. \ref{fig:cubic} and \ref{fig:hex}. Indeed, the spin-splitting properties of altermagnets can be also related to higher-order multipole moments, but of magnetic rather than electric character,~\cite{Song2025, Yershov2024, Aoyama2024, BuiarelliMultipole}. These higher-order magnetic multipoles are a direct consequence of the fact that altermagnets are invariant under a combination of  $\mathcal{T}$ with a symmetry operation that involves rotations, as explained above. Analogously, the higher-order electric multipoles related to the non-Rashba types of SOC effects (i.e., Dresselhaus, Weyl, and Ising) are a manifestation of these systems invariance under a combination of $\mathcal{I}$ and a symmetry operation involving rotations. Recently, Ref. \cite{Visser2026} proposed to identify these types of non-magnetic phases as \textit{alterelectrics}, emphasizing their analogy with altermagnets. In contrast to the latter, however, SOC is necessary for the former to display spin-split bands. In this context, our work provides a full classification of spin-splitting symmetries in the band structure of alterelectrics, including the distinct mechanisms and natures of their corresponding nodal lines.

We note that it is also possible to generate odd-parity spin-splitting without SOC in certain non-collinear coplanar magnetically ordered systems that break $\mathcal{I}$~\cite{Hellenes2023,Comin2025}. The main difference with respect to the cases studied in this paper is that the bands' spins always point to a single direction in the electronic band dispersion, whereas in our case, the expected spins do not remain collinear at different values of $\vb*{k}$.  

The tables and models we present should be beneficial for future studies of SOC-related phenomena in solids, where a general model that applies to more than one space group is needed. In particular, any electronic instability developing in one of these crystalline structures, such as superconductivity, is likely to be strongly impacted by the spin-splitting symmetries. A well-known case is the proposed development of an unusual type of pairing, dubbed Ising-superconductivity~\cite{Ye2015}, in monolayer transition metal dichalcogenides, which display Ising SOC terms. In this pairing state, application of an in-plane magnetic field can promote a nodal topological superconductor, whose properties are derived from the spin-splitting nodal lines~\cite{He2018,Shaffer2020}.  Our analysis opens the door for a systematic analysis of the relationship between spin-splitting nodal structures generated by SOC and exotic pairing properties in NCS superconductors. 
Apart from providing these results and general insight, our study also underlines the usefulness of a symmetry-based approach where irreps that connect lower-symmetry groups to high-symmetry reference groups are used in conjunction with real and reciprocal space Hamiltonians. Applying irrep projection operators onto second quantized operators enables the delineation of different terms allowed in TB models, simplifying calculations and interpretations of electronic structural features.

\section*{Acknowledgements}
F.Y. and T.B. are supported by the NSF CAREER grant DMR-2046020. R.M.F. acknowledges support from the Research Corporation for Science Advancement through the Cottrell SEED Award CS-SEED-2025-012.

\appendix

\renewcommand\thefigure{\thesection\arabic{figure}}   
\renewcommand\thetable{\thesection\arabic{table}}  

\counterwithin{figure}{section}
\counterwithin{table}{section}
\counterwithin{equation}{section}

\section{Character tables of high-symmetry parent point groups}
\label{sm:character}

For reference, we document the group theory conventions used in the main text by providing partial character tables of $\mathcal{I}$-odd, $\mathcal{T}$-even irreps of the two high-symmetry parent point groups $m\bar{3}m$ (Tables~\ref{tab:charactercubic}) and $6/mmm$ (Table~\ref{tab:characterhex}). The symmetry elements listed for each group form a set of generators and are enough to distinguish each irrep and to derive the complete character table. The information is obtained from the Bilbao Crystallographic Server~\cite{Aroyo2006}. 

\begin{table}[!h]
\centering
\caption{Character table of point group $m\bar{3}m$ (generators only, $\mathcal{I}$-odd, $\mathcal{T}$-even irreps).}
\renewcommand{\arraystretch}{1.5}
\begin{tabular}{ 
  | >{\centering\arraybackslash}p{0.12\linewidth}  
  | >{\centering\arraybackslash}p{0.12\linewidth}   
  | >{\centering\arraybackslash}p{0.12\linewidth}   
  | >{\centering\arraybackslash}p{0.12\linewidth}   
  | >{\centering\arraybackslash}p{0.12\linewidth}   
  | >{\centering\arraybackslash}p{0.12\linewidth}   
  | >{\centering\arraybackslash}p{0.12\linewidth}   | }
    \hline 
    Irrep & $\mathcal{E}$ & $4^{+}_{001}$ & $3^{+}_{111}$ & $2_{110}$ & $\mathcal{I}$ & $\mathcal{T}$ \\
    \hline
    $A^+_{1u}$ & 1 & 1 & 1  & 1 & -1 & 1 \\
    \hline
    $A^+_{2u}$ & 1 & -1 & 1  & -1 & -1 & 1 \\
    \hline
    $E^+_{u}$ & 2 & 0 & 2  & 0 & -2 & 2 \\
    \hline
    $T^+_{1u}$ & 3 & 1 & -1  & -1 & -3 & 3 \\
    \hline
    $T^+_{2u}$ & 3 & -1 & -1  & 1 & -3 & 3 \\
    \hline
\end{tabular}
\label{tab:charactercubic}
\end{table}

\begin{table}[!h]
\centering
\caption{Character table of point group $6/mmm$ (generators only, $\mathcal{I}$-odd, $\mathcal{T}$-even irreps).}
\renewcommand{\arraystretch}{1.5}
\begin{tabular}{ 
  | >{\centering\arraybackslash}p{0.12\linewidth}  
  | >{\centering\arraybackslash}p{0.12\linewidth}   
  | >{\centering\arraybackslash}p{0.12\linewidth}   
  | >{\centering\arraybackslash}p{0.12\linewidth}   
  | >{\centering\arraybackslash}p{0.12\linewidth}   
  | >{\centering\arraybackslash}p{0.12\linewidth}   
  | >{\centering\arraybackslash}p{0.12\linewidth}   | }
    \hline 
    Irrep & $\mathcal{E}$ & $6^{+}_{001}$ & $2_{100}$ & $2_{120}$ & $\mathcal{I}$ & $\mathcal{T}$  \\
    \hline
    $A^+_{1u}$ & 1 & 1 & 1 & 1 & -1 & 1 \\
    \hline
    $A^+_{2u}$ & 1 & 1  & -1 & -1 & -1 & 1 \\
    \hline
    $B^+_{1u}$ & 1 & -1  & -1 & 1 & -1 & 1 \\
    \hline
    $B^+_{2u}$ & 1 & -1  & 1 & -1 & -1 & 1 \\
    \hline
    $E^+_{1u}$ & 2 & 1  & 0 & 0 & -2 & 2 \\
    \hline
    $E^+_{2u}$ & 2 & -1 & 0 & 0 & -2 & 2 \\
    \hline
\end{tabular}
\label{tab:characterhex}
\end{table}

\section{Transformation of Rashba Hamiltonian vector in the cubic parent phase}
\label{sm:rashba}

In this section, we show explicitly how the three components of the reciprocal-space Rashba Hamiltonian vector transform under the point group generators of $m\bar{3}m$. We define the three-component pseudo-vector of Rashba Hamiltonian as:
\begin{equation}
\vb*{h}_R \equiv \begin{bmatrix} \mathcal{H}_{R,x} \\ \mathcal{H}_{R,y} \\ \mathcal{H}_{R,z} \end{bmatrix} = \alpha_R \begin{bmatrix} k_y\sigma_z-k_z\sigma_y \\ k_z\sigma_x-k_x\sigma_z \\ k_x\sigma_y-k_y\sigma_x \end{bmatrix},
\end{equation}
where $\vb*{k}$ and $\vb*{\sigma}$ respectively transform as a polar vector ($T_{1u}^-$) and an axial vector ($T_{1g}^-$), i.e., they transform identically under proper rotations, but $\vb*{k}$ takes on a minus sign under inversion $\mathcal{I}$ and $\vb*{\sigma}$ remains unchanged. 
Considering a set of generators of the point group $m\bar{3}m$: $\mathcal{G}=\{4_{001}^+, 3_{111}^+,2_{110},\mathcal{I}\}$ where the crystal direction $[100]$, $[010]$ and $[001]$ are aligned with the $x$, $y$ and $z$-axes, and applying the generators on $\vb*{h}_R$, we obtain: 
\begin{equation}
\label{eqn:H_R_transform_1}
 {4}_{001}^+\begin{bmatrix} \mathcal{H}_{R,x} \\ \mathcal{H}_{R,y} \\ \mathcal{H}_{R,z} \end{bmatrix}= \alpha_R\begin{bmatrix} (-k_x)\sigma_z-k_z(-\sigma_x) \\ k_z\sigma_y-k_y\sigma_z \\ k_y(-\sigma_x)-(-k_x)\sigma_y \end{bmatrix} = \begin{bmatrix} \mathcal{H}_{R,y} \\ -\mathcal{H}_{R,x} \\ \mathcal{H}_{R,z} \end{bmatrix},
\end{equation}
\begin{equation}
 {3}_{111}^+\begin{bmatrix} \mathcal{H}_{R,x} \\ \mathcal{H}_{R,y} \\ \mathcal{H}_{R,z} \end{bmatrix}= \alpha_R\begin{bmatrix} k_z\sigma_x-k_x\sigma_z \\ k_x\sigma_y-k_y\sigma_x \\ k_y\sigma_z-k_z\sigma_y \end{bmatrix} = \begin{bmatrix} \mathcal{H}_{R,y} \\ \mathcal{H}_{R,z} \\ \mathcal{H}_{R,x} \end{bmatrix},
\end{equation}
\begin{equation}
 {2}_{110}\begin{bmatrix} \mathcal{H}_{R,x} \\ \mathcal{H}_{R,y} \\ \mathcal{H}_{R,z} \end{bmatrix}= \alpha_R\begin{bmatrix} k_x(-\sigma_z)-(-k_z)\sigma_x \\ (-k_z)\sigma_y-k_y(-\sigma_z) \\ k_y\sigma_x-k_x\sigma_y \end{bmatrix} = \begin{bmatrix} \mathcal{H}_{R,y} \\ \mathcal{H}_{R,x} \\ -\mathcal{H}_{R,z} \end{bmatrix}, 
\end{equation}
\begin{equation}
\label{eqn:H_R_transform_4}
 \mathcal{I}\begin{bmatrix} \mathcal{H}_{R,x} \\ \mathcal{H}_{R,y} \\ \mathcal{H}_{R,z} \end{bmatrix}= \alpha_R\begin{bmatrix} (-k_y)\sigma_z-(-k_z)\sigma_y \\ (-k_z)\sigma_x-(-k_x)\sigma_z \\ (-k_x)\sigma_y-(-k_y)\sigma_x 
\end{bmatrix} = \begin{bmatrix} -\mathcal{H}_{R,x} \\ -\mathcal{H}_{R,y} \\ -\mathcal{H}_{R,z} \end{bmatrix}.
\end{equation}

In comparison, the electric polarization $\vb*{P}=\begin{bmatrix} p_x \\ p_y \\ p_z \end{bmatrix}$ transforms as a polar vector, in other words, the matrices for point group operations are identical to those that act on cartesian coordinates ($x$,~$y$,~$z$): 
\begin{equation}
 {4}_{001}^+\begin{bmatrix} p_x \\ p_y \\ p_z \end{bmatrix}=\begin{bmatrix} p_y \\ -p_x \\ p_z \end{bmatrix},
\end{equation}
\begin{equation}
{3}_{111}^+\begin{bmatrix} p_x \\ p_y \\ p_z \end{bmatrix}=\begin{bmatrix} p_y \\ p_z \\ p_x \end{bmatrix}, 
\end{equation}
\begin{equation}
{2}_{110}\begin{bmatrix} p_x \\ p_y \\ p_z \end{bmatrix}=\begin{bmatrix} p_y \\ p_x \\ -p_z \end{bmatrix}, \end{equation}
\begin{equation}
{\mathcal{I}}\begin{bmatrix} p_x \\ p_y \\ p_z \end{bmatrix}=\begin{bmatrix} -p_x \\ -p_y \\ -p_z \end{bmatrix},
\end{equation}
which follows an identical pattern as Eq.~\ref{eqn:H_R_transform_1}-\ref{eqn:H_R_transform_4}. 

Other symmetry elements in $m\bar{3}m$ can be generated by combinations of the elements in $\mathcal{G}$, such that we find that $\vb*{h}_R$ transforms identically as $\vb*{P}$ in $m\bar{3}m$.

\section{Explicit expressions of cubic-in-$\vb*{k}$ spin-splitting Hamiltonians as irrep bases of the cubic parent group}
\label{sm:hamiltonian_3rd}

In Sections~\ref{sec:cubic} and \ref{sec:hex}, all linear-in-$\vb*{k}$ SOC terms were tabulated and discussed in detail. In addition to these tersm, other terms that are higher order in  $\vb*{k}$ can be calculated following the same procedure with the projection operators as detailed in the main text. In Table~\ref{table:cubicink}, we present the explicit expressions of cubic-in-$\vb*{k}$ $\mathcal{H}_{SOC}$ bases for each irrep of the high-symmetry parent phase $m\bar{3}m$. All bases are found by applying the corresponding projection operators to appropriate initial guesses, as introduced in Section~\ref{sec:cubic} and \ref{sec:tight binding}. The basis of $A^+_{2u}$, i.e., the D-c effect, is already presented in the last column of Fig.~\ref{fig:cubic}. 

Note that we assigned names these cubic-in-$\vb*{k}$ terms according to their irreps following the same convention as the linear-in-$\vb*{k}$ terms. However, there is no widespread naming convention for these higher order SOC effects, and the spin-patterns of bands under the effect of these effects deviate from those shown in Fig.~\ref{fig:cubic}. 

\begin{table}
\renewcommand{\arraystretch}{1.2}
\centering
\caption{Explicit expressions of cubic-in-$\vb*{k}$ spin-splitting Hamiltonian bases of $\mathcal{I}$-odd irreps of $m\bar{3}m$ (normalization factors omitted, $k^2\equiv k^2_x+k^2_y+k^2_z$).}
\begin{tabular}{ 
  | >{\centering\arraybackslash}p{0.12\linewidth}  
  | >{\centering\arraybackslash}p{0.12\linewidth}   
  | >{\centering\arraybackslash}p{0.12\linewidth}   
  | >{\centering\arraybackslash}p{0.56\linewidth}   | }
\hline
Effect & Irrep & E/ET & $\mathcal{H}_{SOC}(\vb*{k},\vb*{\sigma})$ basis\\
\hline
\multirow{3}{*}[-2em]{R} & \multirow{3}{*}[-2em]{3$T^+_{1u}$} & \multirow{3}{*}[-2em]{E-1} & $\begin{bmatrix} k^3_z\sigma_y-k^3_y\sigma_z \\ k^3_x\sigma_z-k^3_z\sigma_x \\ k^3_y\sigma_x-k^3_x\sigma_y \end{bmatrix}$ \\
\cline{4-4}
& & & $\begin{bmatrix} k^2_x(k_z\sigma_y-k_y\sigma_z) \\ k^2_y(k_x\sigma_z-k_z\sigma_x) \\ k^2_z(k_y\sigma_x-k_x\sigma_y) \end{bmatrix}$ \\
\cline{4-4}
& & & $\begin{bmatrix} k_yk_z(k_y\sigma_y-k_z\sigma_z) \\ k_zk_x(k_z\sigma_z-k_x\sigma_x) \\ k_xk_y(k_x\sigma_x-k_y\sigma_y) \end{bmatrix}$ \\ 
\hline
\multirow{4}{*}[-4em]{D-1} & \multirow{4}{*}[-4em]{4$T^+_{2u}$} & \multirow{4}{*}[-4em]{\parbox{0.048\textwidth}{ET-2 (E-3)}} & $\begin{bmatrix} k^3_z\sigma_y+k^3_y\sigma_z \\ k^3_x\sigma_z+k^3_z\sigma_x \\ k^3_y\sigma_x+k^3_x\sigma_y \end{bmatrix}$ \\
\cline{4-4}
& & & $\begin{bmatrix} k^2_x(k_z\sigma_y+k_y\sigma_z) \\ 
 k^2_y(k_x\sigma_z+k_z\sigma_x) \\ k^2_z(k_y\sigma_x+k_x\sigma_y) \end{bmatrix}$ \\
\cline{4-4}
& & & $\begin{bmatrix} k_yk_z(k_y\sigma_y+k_z\sigma_z) \\ 
k_zk_x(k_z\sigma_z+k_x\sigma_x) \\ k_xk_y(k_x\sigma_x+k_y\sigma_y) \end{bmatrix}$\\
\cline{4-4}
& & & $\begin{bmatrix} k_xk_yk_z\sigma_x \\ k_xk_yk_z\sigma_y \\ k_xk_yk_z\sigma_z \end{bmatrix}$ \\ 
\hline
\multirow{3}{*}[-2em]{D-2} & \multirow{3}{*}[-2em]{3$E^+_{u}$} & \multirow{3}{*}[-2em]{\parbox{0.05\textwidth}{ET-2 (E-5)}} & $\begin{bmatrix} k^3_x\sigma_x-k^3_y\sigma_y \\ -k^3_x\sigma_x-k^3_y\sigma_y+2k^3_z\sigma_z
\end{bmatrix}$ \\
\cline{4-4}
& & & $\begin{bmatrix} k^2(k_x\sigma_x-k_y\sigma_y) \\ k^2(-k_x\sigma_x-k_y\sigma_y+2k_z\sigma_z)
\end{bmatrix}$\\ 
\cline{4-4}
& & & $\begin{bmatrix} (k^2_x-k^2_y)(\vb*{k}\cdot\vb*{\sigma}) \\ (-k^2_x-k^2_y+2k^2_z)(\vb*{k}\cdot\vb*{\sigma}) \end{bmatrix}$\\
\hline
\multirow{2}{*}[-.5em]{W} & \multirow{2}{*}[-.5em]{2$A^+_{1u}$} & \multirow{2}{*}[-.5em]{\parbox{0.05\textwidth}{ET-0 (E-9)}} & $\begin{bmatrix} k^3_x\sigma_x+k^3_y\sigma_y+k^3_z\sigma_z \end{bmatrix}$\\
\cline{4-4}
& & & $\begin{bmatrix} k^2(\vb*{k}\cdot\vb*{\sigma}) \end{bmatrix}$\\
\hline
D-c & 1$A^+_{2u}$ & E-3 & See Fig.~\ref{fig:cubic}.\\
\hline
\end{tabular}
\label{table:cubicink}
\end{table}

\section{DFT calculation of spin patterns in M\lowercase{o}$_3$A\lowercase{l}$_2$C}
\label{sm:dft}

In this section, we present the DFT results of spin patterns of a chiral system, Mo$_3$Al$_2$C (space group $P4_132$), which, according to our analysis in Section~\ref{sec:cubic}, is predicted to host \textit{only} the Weyl SOC terms in its spin-splitting Hamiltonian. The lowest-order-in-$\vb*{k}$ Weyl term goes: $\mathcal{H}_W=\alpha_W\vb*{k}\cdot\vb*{\sigma}$, leading to a hedgehog-like radial spin distribution in each spin-polarized subband around the $\Gamma$ point.

\begin{figure}
\includegraphics[width=\linewidth]{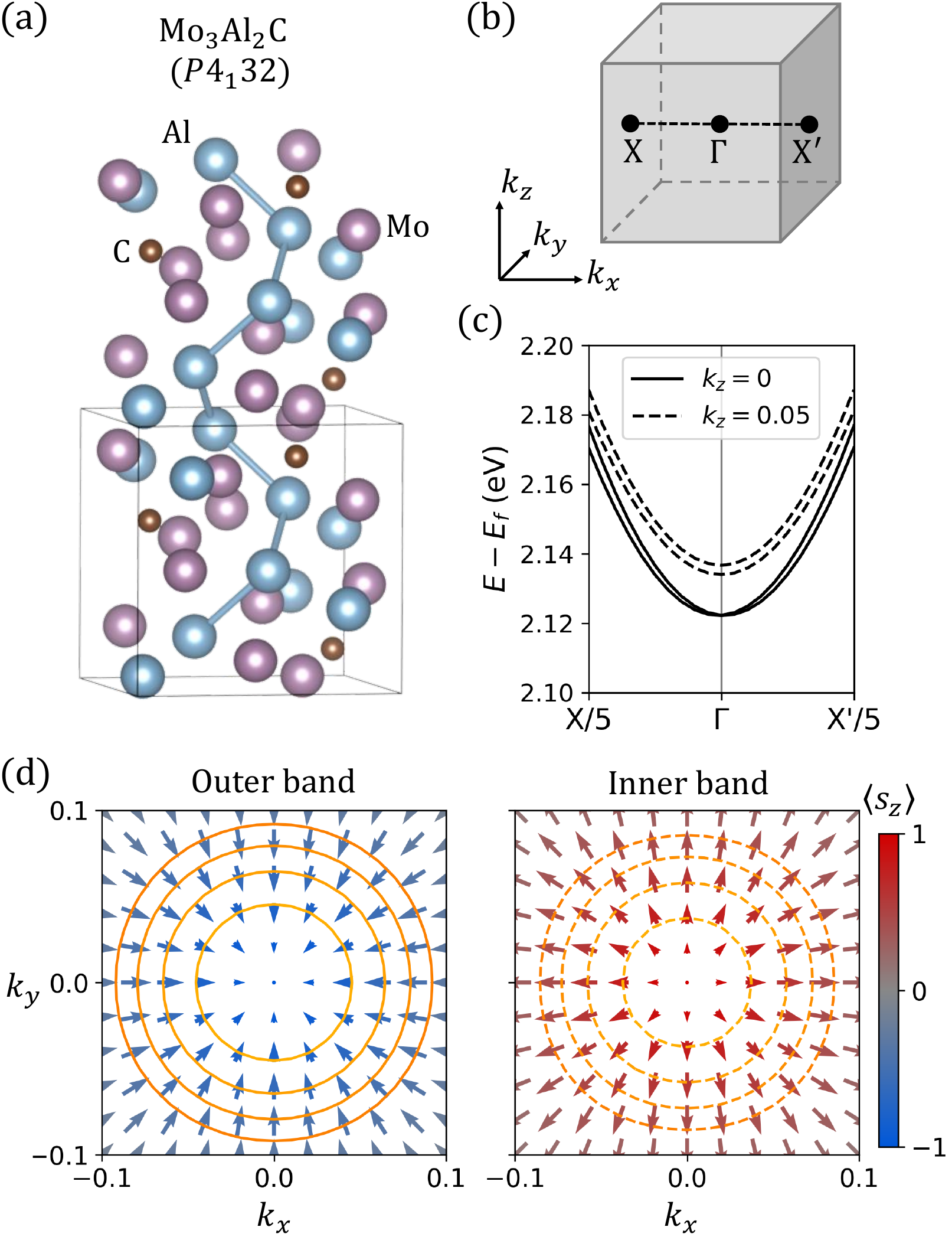}
\caption{(a) Crystal structure of Mo$_3$Al$_2$C. Bonds between Al atoms are partially shown to illustrate the $4_1$ screw axis along $z$. (b) Schematic of the first BZ labeled with the high symmetry path $X'-\Gamma-X$. (c) DFT band structure along $X'/5-\Gamma -X/5$ (solid line) and the path slightly above it at $k_z=0.05$ (dash line). (d) Spin patterns in both subbands at $k_z=0.05$ around the plane center $(0,0,0.05)$. Lengths and directions of the arrows describe the magnitudes and directions of the in-plane spin components at each $\vb*{k}$-point; colors of arrows stand for out-of-plane spin components. Concentric circles represent equi-energy contours with an interval of 10~meV.}
\label{fig:432-dft}
\end{figure}

Mo$_3$Al$_2$C features a non-symmorphic crystal structure where three non-intersecting $4_1$ screw axes lie orthogonal to each other along $x,y,z$, respectively, with one of them shown explicitly in Fig.~\ref{fig:432-dft}(a). DFT calculations of band structures and spin patterns are carried out on both $k_z=0$ and $k_z=0.05$ (unit: $\frac{2\pi}{a}$) planes using VASP~\cite{Kresse1996_DFT1,Kresse1996_DFT2,Kohn1965} with the Projector Augmented Wave (PAW) method~\cite{Blochl1994PAW1,kresse1999PAW2} and the Perdew–Burke–Ernzerhof (PBE) exchange-correlation functional revised for solids (PBEsol)~\cite{Perdew1996_PBE,Perdew2008_PBEsol}. Structural parameters are directly adopted from experimental measurements in Ref.~\cite{Jeitschko1963} without any relaxation. A uniform $8\times8\times8$ $\vb*{k}$-mesh is used for the calculation of charge density, with the cut-off energy set to 600~eV, and the convergence tolerance of self-consistent loops set to $10^{-8}$~eV. SOC is turned on. No magnetic order is included. Expected spin component at each $\vb*{k}$-point along each direction is determined as:
\begin{equation}
\langle s_i(n,\vb*{k}) \rangle = \bra{\psi_{n,\vb*{k}}}\sigma_i\ket{\psi_{n,\vb*{k}}},
\end{equation}
where $\sigma_i$ is the Pauli matrix ($i\in\{x,y,z\}$), and $\ket{\psi_{n,\vb*{k}}}$ is the Kohn-Sham eigenstate of the $n^{th}$ band at $\vb*{k}$ obtained from the DFT calculation. The value of $\langle s_i \rangle$ runs from -1 to 1.

In Fig.~\ref{fig:432-dft}, we present the results of a pair of spin-polarized subbands of Mo$_3$Al$_2$C which are well-isolated from other bands around $\Gamma$, located at $\sim2$~eV above the Fermi level. Excellent agreement is found between DFT results and the theoretical $\mathcal{H}_{SOC}$ constructed in both reciprocal and real spaces shown in Fig.~\ref{fig:cubic}, second from the last column. We plot spin patterns on the $k_x$-$k_y$ plane at $k_z$ slightly larger than 0 to better illustrate the behavior of out-of-plane components (red for spin-up, blue for spin-down) in the wavevector space. 

\section{Tight-binding model for the Dresselhaus-cubic system}
\label{sm:D_c}

\begin{figure}
\includegraphics[width=0.9\linewidth]{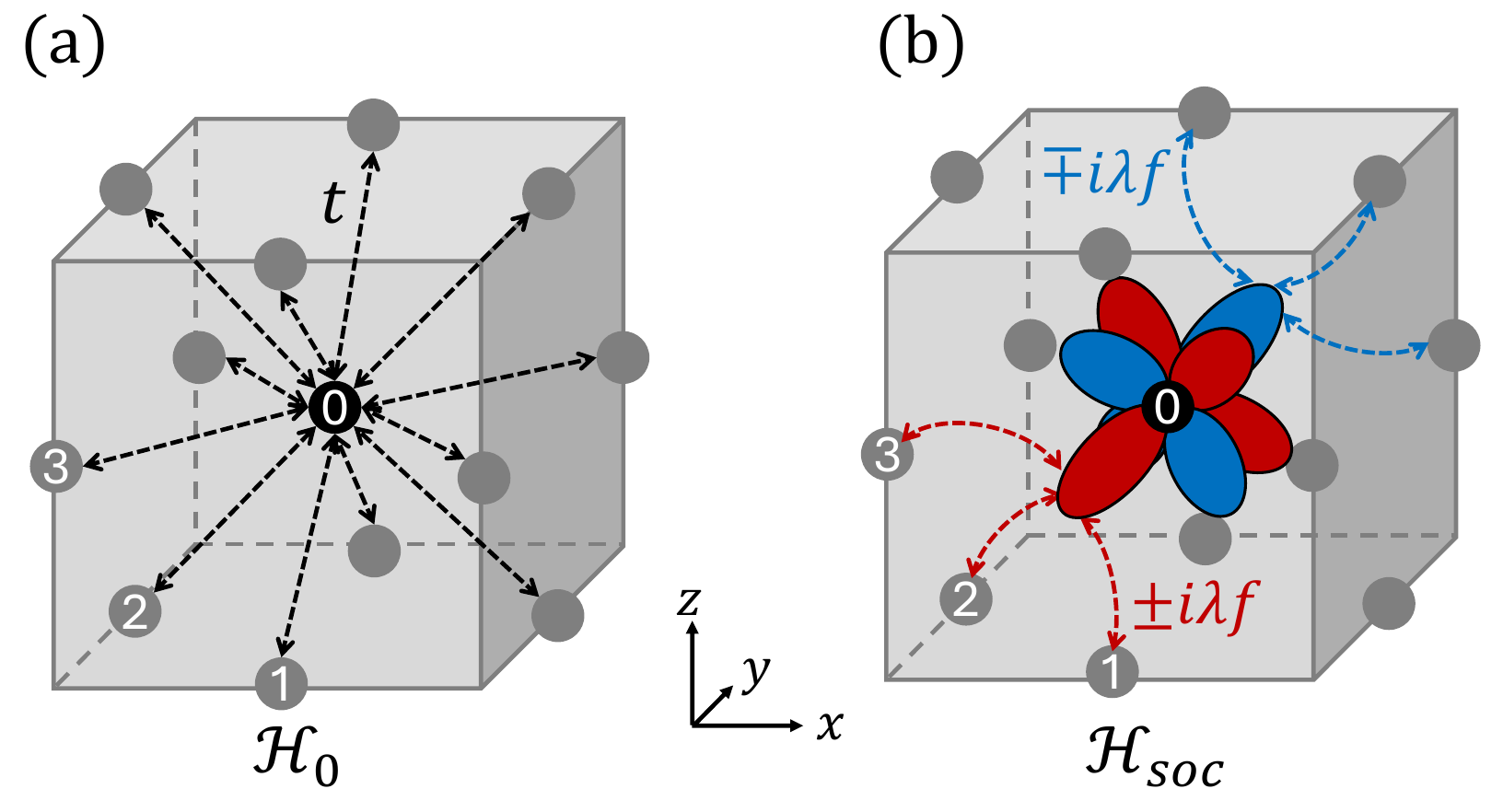}
\caption{Unit cell for and the details of the TB model constructed in the cubic parent phase $m\bar{3}m$ for the SOC term associated with the electric octupole $A_{2u}^+$. Hoppings (dashed arrows) are considered between site 0 and its nearest neighbors (sites 1, 2, 3). (a) The spinless part $\mathcal{H}_{0}$ respects full symmetries of $m\bar{3}m$ and (b) the SOC part $\mathcal{H}_{SOC}$ reproduces the symmetries of the electric multipole present in the system.}
\label{fig:tb_3d}
\end{figure}

The unit cell used in Section~\ref{sec:tight binding} and hoppings therein can not capture all the symmetries broken by the electric octupole $A_{2u}^+$ responsible for the Dresselhaus cubic (D-c) SOC term. To circumvent this problem, we switch to another structure with atom 0 sitting at the body-center of a cubic unit cell (Wyckoff position 1b), and three other atoms (atoms 1, 2, 3) on the edge-centers (Wyckoff position 3d), as illustrated in Fig.~\ref{fig:tb_3d}.

Following the same method and definitions presented in Section~\ref{sec:tight binding}, we enforce the spinless part of the Hamiltonian $\mathcal{H}_0$ to respect all symmetries (including $\mathcal{I}$) of the parent phase $m\bar{3}m$. The SOC part $\mathcal{H}_{SOC}$ has imaginary hoppings that break the same symmetries as those broken by the electric octupole $F=xyz$, which transforms the same way as $\mathcal{H}_{D-c}$ in the point group $m\bar{3}m$.  

With the unit cell in Fig~\ref{fig:tb_3d}, the expression of $\mathcal{H}_0$ is:
\begin{widetext}
\begin{equation}
    \mathcal{H}_0=t\sum_{\vb*{R}}\sum_{i\neq j\neq k} \sum_{\alpha} \left(c_{0\alpha}^{\vb*{R}\dagger} c_{i\alpha}^{\vb*{R}}+c_{0\alpha}^{\vb*{R}\dagger} c_{i\alpha}^{\vb*{R+a_j}}+c_{0\alpha}^{\vb*{R}\dagger} c_{i\alpha}^{\vb*{R+a_k}}+c_{0\alpha}^{\vb*{R}\dagger} c_{i\alpha}^{\vb*{R+a_j+a_k}}\right)+h.c..
\end{equation}
\end{widetext}

For $\mathcal{H}_{SOC}$, the D-c counterpart of Eqn.\ref{eqn:tb_vector} can be written as:
\begin{equation}
\label{eqn:tb_vector_2}
    \vb*{v}^{\vb*{R}}=\begin{bmatrix}
    v_{1,y}^{\vb*{R}}
    \vspace{1mm}\\
    v_{2,z}^{\vb*{R}}
    \vspace{1mm}\\
    v_{3,x}^{\vb*{R}}
    \vspace{1mm}\\
    v_{1,z}^{\vb*{R}}
    \vspace{1mm}\\
    v_{2,x}^{\vb*{R}}
    \vspace{1mm}\\
    v_{3,y}^{\vb*{R}}
    \end{bmatrix}
    =
    \begin{bmatrix}
        \hat{c}_{1}^{\vb*{R}}-\hat{c}_{1}^{\vb*{R+a_2}}+\hat{c}_{1}^{\vb*{R+a_3}}-\hat{c}_{1}^{\vb*{R+a_3+a_2}}
        \vspace{1mm}\\
        \hat{c}_{2}^{\vb*{R}}-\hat{c}_{2}^{\vb*{R+a_3}}+\hat{c}_{2}^{\vb*{R+a_1}}-\hat{c}_{2}^{\vb*{R+a_1+a_3}}
        \vspace{1mm}\\
        \hat{c}_{3}^{\vb*{R}}-\hat{c}_{3}^{\vb*{R+a_1}}+\hat{c}_{3}^{\vb*{R+a_2}}-\hat{c}_{3}^{\vb*{R+a_2+a_1}}
        \vspace{1mm}\\

        \hat{c}_{1}^{\vb*{R}}-\hat{c}_{1}^{\vb*{R+a_3}}+\hat{c}_{1}^{\vb*{R+a_2}}-\hat{c}_{1}^{\vb*{R+a_2+a_3}}
        \vspace{1mm}\\
        \hat{c}_{2}^{\vb*{R}}-\hat{c}_{2}^{\vb*{R+a_1}}+\hat{c}_{2}^{\vb*{R+a_3}}-\hat{c}_{2}^{\vb*{R+a_3+a_1}}
        \vspace{1mm}\\
        \hat{c}_{3}^{\vb*{R}}-\hat{c}_{3}^{\vb*{R+a_2}}+\hat{c}_{3}^{\vb*{R+a_1}}-\hat{c}_{3}^{\vb*{R+a_1+a_2}}
    \end{bmatrix}.
\end{equation}

By applying ~the projection operator of $A_{2u}^+$, we get the form of the $M$ coefficient matrix for the D-c term as:
\begin{equation}
    M=\begin{bmatrix} 
    0 & 0 & f & 0 & -f & 0 \\
    f & 0 & 0 & 0 & 0 & -f \\
    0 & f & 0 & -f & 0 & 0
    \end{bmatrix}.
\end{equation}

The spin pattern obtained from this minimal TB model is presented in the main text in Fig.~\ref{fig:cubic}(c), last column.

\section{Tight-binding models for spin-splitting multipoles in NCS subgroups of the hexagonal group $6/mmm$}
\label{sm:tight binding_hex}

\begin{figure}
\includegraphics[width=0.9\linewidth]{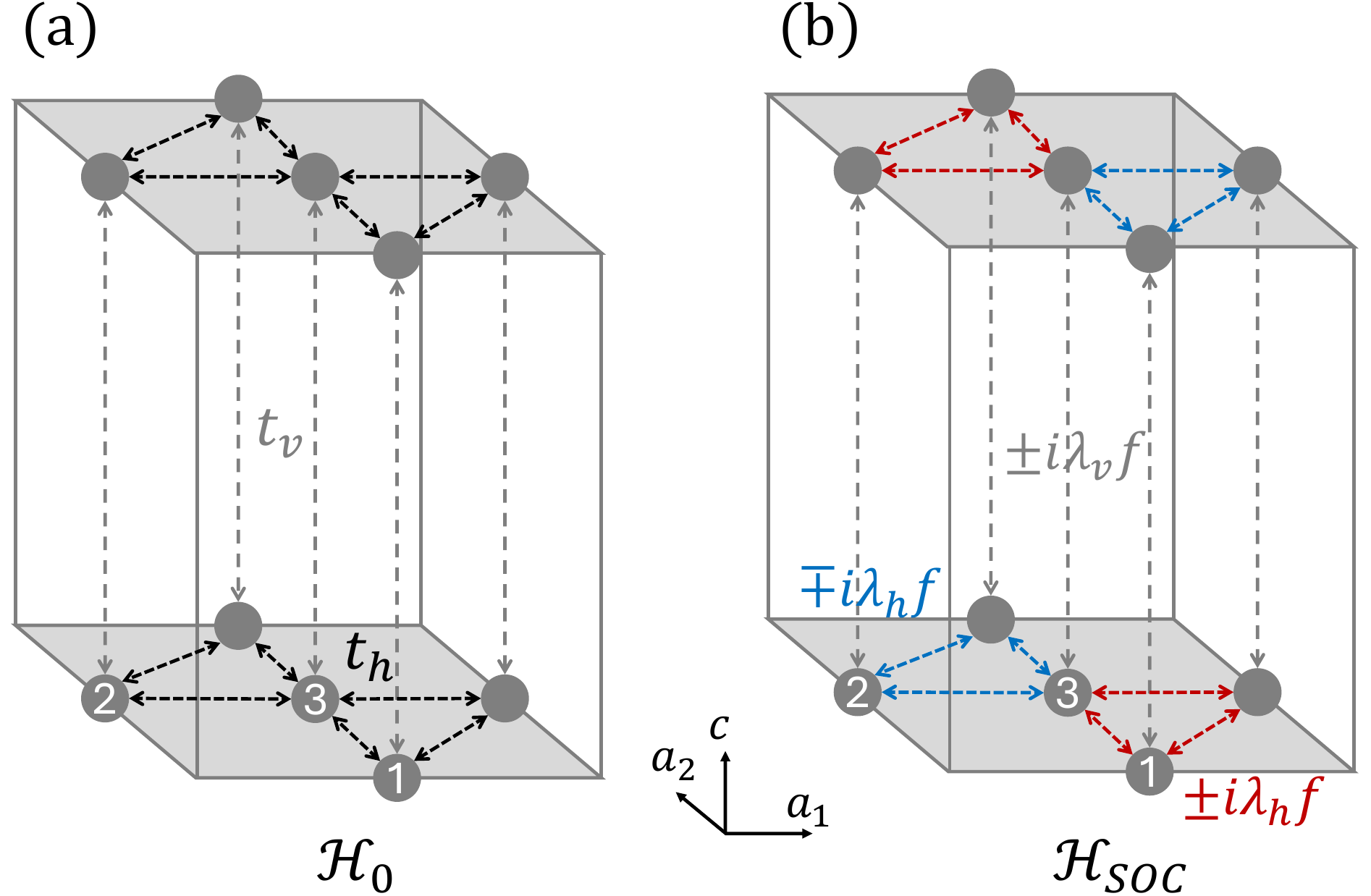}
\caption{Example of a TB model constructed in the hexagonal parent phase $6/mmm$ corresponding to an $\mathcal{I}$-breaking SOC term. The $a_1$-axis is aligned with the $x$-axis. Hoppings (dashed lines) are considered among three kagome sites (1, 2, 3) between both in-plane ($t_h$, $\lambda_h$) and out-of-plane ($t_v$, $\lambda_v$) nearest neighbors. (a) The spinless part $\mathcal{H}_{0}$ respects full symmetries of $6/mmm$ and (b) the SOC part $\mathcal{H}_{SOC}$ reproduces the symmetries of the electric multipole present in the system.}
\label{fig:tb_hex}
\end{figure}

In this appendix, we construct effective TB models to reproduce spin-splitting effects arising from $\mathcal{I}$-breaking irreps in the hexagonal parent group $P6/mmm$ following the same method introduced in Section~\ref{sec:tight binding} for the cubic parent phase $Pm\bar{3}m$.

We consider an $s$-like orbital on each atom located at the Wyckoff position 3f of the space group $P6/mmm$ (Fig.~\ref{fig:tb_hex}). This model represents the kagome structure in simple stacking. Using the definition in Eqn.~\ref{ac_vec}, we define two basis vectors of annihilation operator vectors:
\begin{align}
\vb*{c}^{\vb*{R}}_h & =(\hat{c}_1^{\vb*{R}},\hat{c}_1^{\vb*{R+a_2}},\hat{c}_2^{\vb*{R}},\hat{c}_2^{\vb*{R+a_1}},\hat{c}_3^{\vb*{R}})^T, \\
\vb*{c}^{\vb*{R}}_v & =(\hat{c}_1^{\vb*{R}},\hat{c}_1^{\vb*{R+c}},\hat{c}_2^{\vb*{R}},\hat{c}_2^{\vb*{R+c}},\hat{c}_3^{\vb*{R}},\hat{c}_3^{\vb*{R+c}})^T.
\end{align}
Considering only hoppings between in-plane and out-of-plane nearest neighbors, we can write the general forms of the spinless ($\mathcal{H}_0$) and spin-splitting ($\mathcal{H}_{SOC}$) Hamiltonians as:
\begin{align}
\mathcal{H}_0 & =\sum_{\vb*{R}}\Big(t_h\vb*{\hat{c}}^{\vb*{R}\dagger}_h M^h_0\vb*{\hat{c}}^{\vb*{R}}_h +t_v\vb*{\hat{c}}^{\vb*{R}\dagger}_v M^v_0\vb*{\hat{c}}^{\vb*{R}}_v\Big),\\
\mathcal{H}^{\Gamma}_{SOC} & =i\sum_{\vb*{R}}\Big(\lambda_h\vb*{\hat{c}}^{\vb*{R}\dagger}_h M^{\Gamma,h}_{SOC}\vb*{\hat{c}}^{\vb*{R}}_h +\lambda_v\vb*{\hat{c}}^{\vb*{R}\dagger}_v M^{\Gamma,v}_{SOC}\vb*{\hat{c}}^{\vb*{R}}_v\Big).
\end{align} 
Here, $M$'s are coefficient matrices to be determined by the symmetries of the system, with $M^h$ accounting for in-plane hoppings and $M^v$ accounting for out-of-plane ones, which are decoupled in hexagonal systems due to the anisotropy along $\vb*{c}$. The spin-independent part $\mathcal{H}_0$ respects all symmetries of $P6/mmm$, such that:
\begin{align}
M^h_0 & =\begin{bmatrix}
0 & 0 & 0 & \sigma_0 & \sigma_0 \\
0 & 0 & \sigma_0 & 0 & \sigma_0 \\
0 & \sigma_0 & 0 & 0 & \sigma_0 \\
\sigma_0 & 0 & 0 & 0 & \sigma_0 \\
\sigma_0 & \sigma_0 & \sigma_0 & \sigma_0 & 0
\end{bmatrix},\\
M^v_0 & =\begin{bmatrix}
0 & \sigma_0 & 0 & 0 & 0 & 0\\
\sigma_0 & 0 & 0 & 0 & 0 & 0 \\
0 & 0 & 0 & \sigma_0 & 0 & 0 \\
0 & 0 & \sigma_0 & 0 & 0 & 0 \\
0 & 0 & 0 & 0 & 0 & \sigma_0 \\
0 & 0 & 0 & 0 & \sigma_0 & 0 
\end{bmatrix},
\end{align}
where $\sigma_0$ is the $2\times2$ identity matrix, and $0$ should be interpreted as the $2\times2$ zero matrix.

The spin-splitting part $\mathcal{H}^{\Gamma}_{SOC}$ breaks the symmetries the corresponding irrep $\Gamma$ breaks, and thus takes a unique form for each irrep. In Table~\ref{table:tb_hex}, we list the non-zero components of $M^{\Gamma,h}_{SOC}$ and $M^{\Gamma,v}_{SOC}$ for each relevant irrep. While we do not list the results for two-dimensional irreps (i.e., $E_{1u}^+$ and $E_{2u}^+$) explicitly in the manuscript, they can be found in the data repository as python codes along with the other TB models \cite{DataRepository}. The spin-splitting effects of these irreps are only allowed in orthorhombic or lower-symmetry systems, making them more intuitive to approach from the cubic parent phase $Pm\bar{3}m$ with simpler expressions. The corresponding spin patterns in the lowest spin-polarized band in each case are listed in Fig.~\ref{fig:hex}(c).

In the current 3-atom kagome model (Fig.~\ref{fig:tb_hex}), the SOC-induced hoppings that transform according to the irrep $B_{2u}^+$ are forbidden among nearest neighbors on the same $x-y$ plane. To circumvent this problem, we work with the second-nearest-neighbor in-plane hoppings using a different vector of operators in this case: 
\begin{equation}
\vb*{c'}^{\vb*{R}}_h=(\hat{c}_1^{\vb*{R}},\hat{c}_1^{\vb*{R-a_1}},\hat{c}_2^{\vb*{R}},\hat{c}_2^{\vb*{R-a_2}},\hat{c}_3^{\vb*{R}},\hat{c}_3^{\vb*{R-a_1-a_2}})^T,
\end{equation}
and substituting $M^h_{SOC}$ with ${M'}^h_{SOC}$, whose non-zero components are listed in the second from the last row of Table~\ref{table:tb_hex}.

\begin{table}
\renewcommand{\arraystretch}{1.4}
\centering
\caption{Coefficient matrices of TB models for spin-splitting effects associated with $\mathcal{I}$-breaking, $\mathcal{T}$-invariant irreps of the hexagonal parent phase $6/mmm$. Only nonzero elements of the $M$ matrices are listed for brevity.}
\begin{tabular}{ 
  | >{\centering\arraybackslash}p{0.1\linewidth}  
  | >{\centering\arraybackslash}p{0.1\linewidth}   
  | >{\centering\arraybackslash}p{0.13\linewidth}      
  | >{\centering\arraybackslash}p{0.59\linewidth}   | }
\hline
Effect & Irrep & Matrix & Non-zero components ($M_{ij}=-M_{ji}$)\\
\hline
\multirow{2}{*}[-1.5em]{R} & \multirow{2}{*}[-1.5em]{1$A^+_{2u}$} & \multirow{1}{*}[-1em]{$M^h_{SOC}$} & $M_{14}=-M_{23}=f(\frac{\sqrt{3}}{2}\sigma_x-\frac{1}{2}\sigma_y)$, $M_{51}=-M_{52}=f(\frac{\sqrt{3}}{2}\sigma_x+\frac{1}{2}\sigma_y)$, $M_{45}=-M_{35}=-f\sigma_y$\\
\cline{3-4}
& & $M_{SOC}^{v}$ & None \\
\hline
\multirow{2}{*}[-1.5em]{W} & \multirow{2}{*}[-1.5em]{2$A^+_{1u}$} & \multirow{1}{*}[-1em]{$M_{SOC}^{h}$} & $M_{14}=-M_{23}=f(\frac{1}{2}\sigma_x+\frac{\sqrt{3}}{2}\sigma_y)$, $M_{51}=-M_{52}=f(\frac{1}{2}\sigma_x-\frac{\sqrt{3}}{2}\sigma_y)$, $M_{45}=-M_{35}=-f\sigma_x$ \\
\cline{3-4}
& & $M_{SOC}^{v}$ & $M_{12}=M_{34}=M_{56}=f\sigma_z$ \\
\hline
\multirow{2}{*}[-2em]{I-1} & \multirow{2}{*}[-2em]{2$B^+_{1u}$} & \multirow{1}{*}[-0.5em]{$M^h_{SOC}$} & $M_{14}=M_{51}=M_{45}=f\sigma_z$,
$M_{23}=M_{52}=M_{35}=-f\sigma_z$\\
\cline{3-4}
& & \multirow{1}{*}[-1em]{$M^v_{SOC}$} & $M_{12}=f\sigma_x$, $M_{34}=f(\frac{1}{2}\sigma_x-\frac{\sqrt{3}}{2}\sigma_y)$, $M_{56}=-f(\frac{1}{2}\sigma_x+\frac{\sqrt{3}}{2}\sigma_y)$\\
\hline
\multirow{2}{*}[-2em]{I-2} & \multirow{2}{*}[-2em]{2$B^+_{2u}$} & \multirow{1}{*}[-0.5em]{${M'}^h_{SOC}$} & $M_{24}=M_{52}=M_{45}=f\sigma_z$,
$M_{13}=M_{61}=M_{36}=-f\sigma_z$ \\
\cline{3-4}
& & \multirow{1}{*}[-1em]{$M^v_{SOC}$} & $M_{12}=f\sigma_y$, $M_{34}=-f(\frac{\sqrt{3}}{2}\sigma_x+\frac{1}{2}\sigma_y)$, $M_{56}=-f(\frac{\sqrt{3}}{2}\sigma_x-\frac{1}{2}\sigma_y)$\\ 
\hline
\end{tabular}
\label{table:tb_hex}
\end{table}

\section{Secondary order parameters and nodal features induced by multiple primary order parameters}
\label{sm:nodal}

In Section~\ref{sec:secondary}, we listed different secondary OPs and nodal structures induced by each primary OP in the high-symmetry parent groups $m\bar{3}m$ (Table~\ref{table:nodes}) and $6/mmm$ (Table~\ref{table:nodes_hex}). However, there are some NCS subgroups of $m\bar{3}m$ and $6/mmm$ that \textit{cannot} be reached by a single OP (e.g., $\bar{6}$ and $\bar{4}$). In these cases, a single primary order parameter cannot connect the low-symmetry group to the reference high-symmetry one. Instead, two or more irreps of the parent space group that do not induce each other need to be considered simultaneously. Even though neither of these irreducible representations is primary by itself, the combination of these two irreducible representations can still lead to secondary order parameters. 

\begin{table}
\caption{Secondary OPs induced by primary OPs of the $m\bar{3}m$ parent group and the corresponding nodal structures in the vicinity of $\Gamma$.}
\renewcommand{\arraystretch}{1.5}
\begin{tabular}{ 
  | >{\centering\arraybackslash}p{0.10\linewidth}  
  | >{\centering\arraybackslash}p{0.29\linewidth}   
  | >{\centering\arraybackslash}p{0.22\linewidth}  
  | >{\centering\arraybackslash}p{0.11\linewidth} 
  | >{\centering\arraybackslash}p{0.18\linewidth} | }
\hline
\multicolumn{2}{| >{\centering\arraybackslash}p{0.39\linewidth}|}{Primary OP} & Secondary OP & Sub-group & Nodal structure\\
\hline
D & {$T_{2u}^+$(0,0,a) $\oplus$ $E_{u}^+$(a,0)/$A_{2u}^+$(a)} & {$A_{2u}^+$(a)/ $E_{u}^+$(a,0)} & {$\bar{4}$} & {[001] line} \\ 
\hline
R-W & {$T_{1u}^+$(0,0,a) $\oplus$ $A_{1u}^+$(a)} & none & {$4$} & {point} \\ 
\hline
R-W & {$T_{1u}^+$(a,a,a) $\oplus$ $A_{1u}^+$/$A_{2u}^+$(a)} & {$A_{2u}^+$/$A_{1u}^+$(a)} & {$3$} & {point} \\ 
\hline    
\end{tabular}
\label{table:doubleirrep_cubic}
\end{table}

In Tables~\ref{table:doubleirrep_cubic} and \ref{table:doubleirrep_hex}, we present detailed information on point groups that can be reached only via two or more irreps, including the secondary irreps induced, as well as nodal features. The results in these tables further support the statement that every non-chiral NCS point group hosts nodal lines.

\begin{table} 
\caption{Secondary OPs induced by primary OPs of the $6/mmm$ parent group and the corresponding nodal structures in the vicinity of $\Gamma$.}
\renewcommand{\arraystretch}{1.5}
\begin{tabular}{ 
  | >{\centering\arraybackslash}p{0.10\linewidth}  
  | >{\centering\arraybackslash}p{0.29\linewidth}   
  | >{\centering\arraybackslash}p{0.22\linewidth}  
  | >{\centering\arraybackslash}p{0.11\linewidth} 
  | >{\centering\arraybackslash}p{0.18\linewidth} | }
  \hline
\multicolumn{2}{| >{\centering\arraybackslash}p{0.39\linewidth}|}{Primary OP} & Secondary OP & Sub-group & Nodal structure\\
    \hline
    {I} &{$B_{1u}^+$(a) $\oplus$ $B_{2u}^+$(a) } & {none} & {$\bar{6}$} & {[001] line, lines$^*$} \\ 
    \hline
    {W-R} &{$A_{1u}^+$(a) $\oplus$ $A_{2u}^+$(a)} & {none} & {$6$} & {point} \\ 
    \hline
    {W-I} &{$A_{1u}^+$(a) $\oplus$ $B_{1u}^+$/$B_{2u}^+$(a) } & {none} & {$32$} & {point} \\ 
    \hline
    {R-I} &{$A_{2u}^+$(a) $\oplus$ $B_{1u}^+$/$B_{2u}^+$(a) } & {none} & {$3m$} & {[001] line} \\ 
    \hline
    {I-W/R} &{$B_{1u}^+$(a) $\oplus$ $B_{2u}^+$(a) $\oplus$ $A_{1u}^+$/$A_{2u}^+$(a) } & {$A_{2u}^+$/$A_{1u}^+$(a)} & {$3$} & {point} \\ 
    \hline
\end{tabular}
\raggedright
\footnotesize{$^*$ Accidental degeneracy}\\
\label{table:doubleirrep_hex}
\end{table}

\begin{figure}
\includegraphics[width=\linewidth]{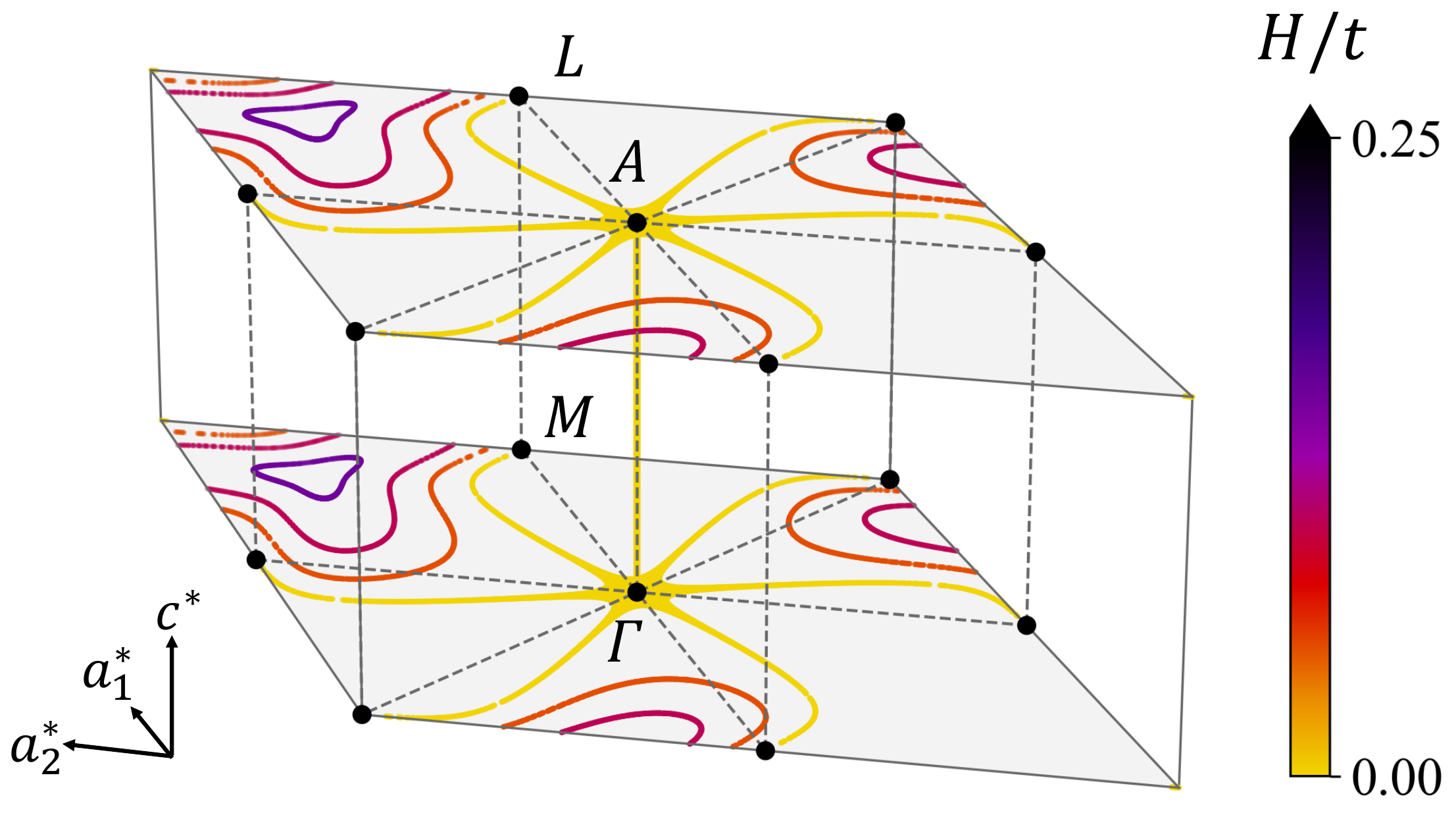}
\caption{Evolution of the nodal structure as a function of Zeeman splitting $H$ (indicated by the color scheme) in the two lowest bands of the TB model with SOC terms corresponding to the combined condensation of $B_{1u}^+$ and $B_{2u}^+$ from the hexagonal parent phase $6/mmm$. At $H=0$ (yellow), there are three nodal lines on each mirror-invariant plane, $k_z=0$ and $k_z=\pi/c$, crossing with each other at $\Gamma$ and $A$, respectively. As $H$ increases, the lines merge to form triangular-shape nodal loops, and finally disappear. The nodal line along $\Gamma-A$ is only present when $H=0$.}
\label{fig:nodal_hex}
\end{figure}

Specifically, in Fig.~\ref{fig:nodal_hex}, we take the point group $\bar{6}$ (reached by the simultaneous condensation of $B_{1u}^{+}$ and $B_{2u}^{+}$ from the hexagonal parent phase $6/mmm$) and illustrate the evolution of the three symmetry-imposed accidental nodal lines on each of its mirror-invariant plane under an external magnetic field along $\vb*{z}$, which shares a similar behavior as in the point group $m$ as illustrated in Fig.~\ref{fig:nodal}.

\clearpage

\end{document}